\documentclass[pre,twocolumn,showpacs,showkeys,preprintnumbers,amsmath,amssymb]{revtex4}

\usepackage{graphicx}
\usepackage{dcolumn}
\usepackage{bm}
\usepackage{amsmath}
\usepackage{amsfonts}
\usepackage{amssymb}
\usepackage{amsthm}
\usepackage{newlfont}
\usepackage{graphicx}
\usepackage{amscd}

\begin{document}


\title{Finite Random Domino Automaton
}

\author{Mariusz Bia{\l}ecki}
\email{bialecki@igf.edu.pl}
 
\affiliation{%
Institute of Geophysics, Polish Academy of Sciences \\
ul.~Ks. Janusza 64, 01-452 Warszawa, Poland}%

\date{\today}

\begin{abstract}
Finite version of Random Domino Automaton (FRDA) - recently proposed in \cite{BiaCzAA} as a toy model of earthquakes - is investigated. 
Respective set of equations describing stationary state of the FRDA is derived and compared with infinite case. 
It is shown that for the system of big size, these equations are coincident with RDA equations.
We demonstrate a non-existence of exact equations for size $N\geq5$ and propose appropriate approximations, the quality of which 
is studied in examples obtained within Markov chains framework.

We derive several exact formulas describing properties of the automaton, including time aspects. 
In particular, a way to achieve a quasi-periodic like behaviour of RDA  is presented. 
Thus, based on the same microscopic rule - which produces exponential and inverse-power like distributions - we extend applicability of the model to quasi-periodic phenomena. 
\end{abstract}


\maketitle


\section{Introduction}

The Random Domino Automaton, proposed in \cite{BiaCzAA}, is a stochastic cellular automaton with avalanches. 
It was introduced as a toy model of earthquakes, but  can be also regarded as an substantial extension of 
1-D forest-fire model proposed by Drossel and Schwabl \cite{DSfire,DCS1Dfire,MMTfire}.

The remarkable feature of the RDA  is the explicit one-to-one relation between details of the dynamical rules of the automaton
(represented by rebound parameters $\mu_i/\nu$ defined in cited article and also below) 
and the produced stationary distribution $n_i$ of clusters of size $i$, which implies distribution of avalanches. 
It is already shown how to reconstruct details of the "microscopic" dynamics from the observed
"macroscopic" behaviour of the system  \cite{BiaCzAA, BiaFRDA1to1}.

As a field of application of RDA we studied a possibility of constructing the Ito equation from a given time series and - in a broader sense - applicability of Ito equation as a model of natural phenomena. 
For RDA - which plays a role of a fully controlled stochastic natural phenomenon - the relevant Ito equation can be constructed in two ways: derived directly from equations and by histogram method from generated time series. 
Then these two results are compared and investigated in \cite{CzBiaTL, CzBiaAG}.

Note that the set of equations of the RDA in a special limit case reduces to the recurrence, which 
leads to known integer sequence - the Motzkin numbers, which establishes a new, remarkable link between the combinatorial object and the stochastic cellular automaton \cite{BiaMotzkin}. 

In the present paper a finite version of Random Domino Automaton is investigated. The mathematical formulation in finite case is 
precise and the presented results clarify which formulas are exact and allow to estimate approximations we impose in infinite case
presented in \cite{BiaCzAA}.
We also show, that equations of finite RDA can reproduce results of \cite{BiaCzAA}, when size $N$ of the system is increasing and 
distributions satisfy an additional assumption ($ n_i \rightarrow 0$ for big $i$). 

On the other hand, a time evolution of Finite RDA can exhibit a periodic-like behaviour (the assumption $ n_i \rightarrow 0$ for big $i$ is violated), which is a novel property. 
Thus, based on the same microscopic rules, depending on a choice of parameters of the model, a wide range of 
properties is possible to obtain.
In particular, such behaviour is interesting in the context of recurrence parameters of earthquakes (see e.g. \cite{WeathRecCA,Parsons}). 
For other simple periodic-like models, see \cite{PachMin, Pach08}. 

The finite case makes an opportunity to employ Markov chains techniques to analyse RDA. Investigating the automaton in 
Markov chains framework we arrive at several novel conclusions, in particular related to expected waiting times for 
some specified behaviour. 

This article completes and substantially extends previous studies of RDA on the level of mathematical structure.
We analyse properties of the automaton related to time evolution and others, as a 
preparation for further prospective comparisons with natural phenomena, including earthquakes.
A matter of adjusting the model to the real data is left for the forthcoming paper.

The plan of the article is as follows. Mimicking \cite{BiaCzAA} in Section \ref{sec:def} we define the finite RDA. 
In Section \ref{sec:equations} we derive respective equations for finite RDA.
In  Section \ref{sec:cases} we will specify them for some chosen cases.
In Section \ref{sec:Markov} we will shortly describe Markov chains setting and describe time aspects of FRDA.
Several examples are presented in Section \ref{sec:examples}.
The last Section \ref{sec:conclusions} contain conclusions and remarks. 
In the Appendix we show non existence of exact equations for RDA as well as present supplementary formulas and Table \ref{tab:statesN10} displaying all states of RDA of size $N=10$.

\section{Finite RDA}
\label{sec:def}

The rules for Finite Random Domino Automaton are the same as in \cite{BiaCzAA}.
We assume: \\
- space is 1-dimensional and discrete -- consists of $N$ cells;\\
- periodic boundary conditions (the last cell is adjacent to the first one);\\
- cell may be in one of two states: empty or occupied by a single ball;\\
- time is discrete and in each time step an incoming ball hits one arbitrarily chosen cell (each cell is equally probable).
      
The state of the automaton changes according to the following rule:\\
$\bullet$ if the chosen cell is empty it becomes occupied with probability $\nu$; with probability $(1-\nu)$ the incoming ball is rebounded and the state remains unchanged; \\
$\bullet$ if the chosen cell is occupied, the incoming ball provokes an avalanche with probability $\mu$ (it removes balls from hit cell and from all adjacent cells); with probability $(1-\mu)$ the incoming ball is rebounded and the state remains unchanged.

The parameter $\nu$ is assumed to be a constant but the parameter $\mu$ is allowed to be a function of size of the hit cluster.
The way in which the probability of removing a cluster depends on its size strongly influences evolution of the system and leads
to various interesting properties, as presented in the following sections. We note in advance that in fact there is only one 
effective parameter $\mu/\nu$ which affects properties of the automaton. Changing of $\mu$ and $\nu$ proportionally 
in a sense corresponds to a rescaling of time unit. 

A diagram shown below presents an automaton of size $N=12$, with three clusters (of size $1,2$ and $4$) in time $t$. An incoming ball
provokes an relaxation of the size {\bf two}, thus in time $t+1$ there are two clusters (of size $1$ and $4$).
\begin{center} 
		\begin{tabular}{l c|c|c|c|c|c|c|c|c|c|c|c|c|c}
		\multicolumn{9}{c}{ \ } &  \multicolumn{1}{c}{$  \stackrel{\boldsymbol{\downarrow}}{\bullet} $}&  \multicolumn{5}{c}{\ }		\\
		 time $= t $   &  $\quad \quad \hookrightarrow$ & $ \bullet $ & $ \bullet $ & $ \ $ & $ \ $ & $ \bullet $ & $ \ $ & $\bullet$ &
			$ {\bullet} $ & $ \ $ & $ \ $ & $ \bullet $ & $ \bullet $ &$ \hookleftarrow \quad \quad $	   \\	
\multicolumn{15}{c}{ \ }		\\
		 time $= t + 1 $  &  $ \quad \quad \hookrightarrow$ & $\bullet$ & $\bullet$ & $ \  $ & $ \ $ & $ \bullet $ & $ \ $ &
$\boldsymbol\downarrow $ & $\boldsymbol\downarrow$ &  $ \ $ & $ \ $ & $\bullet$ & $ \bullet $ & $\hookleftarrow \quad \quad$    \\
\multicolumn{8}{c}{ \ } &  \multicolumn{1}{c}{$ {\bullet} $} & 
\multicolumn{1}{c}{$ {\bullet} $} &  \multicolumn{5}{c}{\ }		\\
			\multicolumn{15}{c}{ \ }		\\
		\end{tabular} 
\end{center}	

Denote by $n_i, i = 1, \ldots, N$ the number of clusters of length $i$, and by 
$n_i^0, i = 1, \ldots, N$ the number of empty clusters of length $i$.
Due to periodic boundary conditions, the number of clusters is equal to the number of empty clusters in the lattice if 
two cases are excluded - when the lattice is full (single cluster of size $N$) and when the lattice is empty (single empty cluster of size $N$). Hence for
\begin{equation}
n_R = \sum_{i=1}^{N-1} n_i, \quad \text{and} \quad  n_R^0= \sum_{i=1}^{N-1} n_i^0
\label{eq:nR}
\end{equation} 
we have
\begin{equation}
n_R =n_R^0.
\label{eq:nn0}
\end{equation}
The density $\rho$ of the system is defined as 
\begin{equation}
\rho = \frac{1}{N}\sum_{i=1}^N  n_i i.
\label{eq:def_rho}
\end{equation}
In this article we investigate a stationary state of the automaton and hence the variables $n_i, n_R, \rho$ and others 
are expected values and do not depend of time.

\section{Equations for finite RDA} \label{sec:equations}

In this section we derive equations describing stationary state of finite RDA. The general idea of the reasoning presented below is: the gain and loss terms balance one another.

\subsection{Balance of density $\rho$}

The density $\rho$ may increase only if an empty cell becomes occupied, and the gain per one time step is $1/N$. It happens with probability $\sim \nu (1-\rho)$. Density losses are realized by avalanches and may be of various size. The effective loss is a product of the size $i$ of the avalanche and probability of its appearance $ \mu_i (n_i i)/N $. Any size $i$ contribute, hence the 
balance of $\rho$ reads  
\begin{equation}
\nu(1-\rho)= \frac{1}{N} \sum_{i=1}^N \mu_i n_i i^2.
\label{eq:rho_balance}
\end{equation}  
We emphasise, the above result is exact -- no correlations were neglected. 
Its form is directly analogous to the respective formula in \cite{BiaCzAA}.

\subsection{Balance of the total number of clusters }

{\bf Gain.} A new cluster (can be of size $1$ only) can be created in the interior of empty cluster of size $\geq 3$. 
$$ \hookrightarrow \cdots | \bullet | \underbrace{ \quad \ | \overbrace{\quad | \cdots \cdots| \quad }^{(i-2) \ \text{cells} \ = \ \text{interior}} | \ \quad }_{i} | \bullet | \cdots \hookleftarrow $$
If the empty cluster is of size $N$, then each cell is in interior. Summing up contributions for all empty clusters, 
the probability is 
\begin{equation}
\sim \sum_{i=3}^{N-1}\nu \left( \frac{i-2}{i}\right) \frac{n_i^0 i }{N} + \nu n_N^0,
\label{eq:ni_gains_p}
\end{equation}
which can take a form (for $N\geq3$)
\begin{equation}
\sim \nu(1-\rho) -2\nu \frac{n_R}{N} + \nu \frac{n_1^0}{N}.
\label{eq:ni_gains}
\end{equation}

{\bf Loss.} Two ways contribute: joining a cluster with another one and removing a cluster due to avalanche.

Joining of two clusters can occur if there exists an empty cluster of length $1$ between them. The exception is when the empty $1$-cluster is the only one empty cluster, and the system consists of a single cluster of length $N-1$. 
Hence, the probability of joining  two clusters is 
\begin{equation}
\sim \nu \left( \frac{n_1^0}{N} - n_{N-1}\right).
\label{eq:balnjoin}
\end{equation}
The probability of avalanche is just
\begin{equation}
\sim \sum_{i=1}^N \mu_i \frac{n_i i}{N}.
\label{eq:balnav}
\end{equation}

Gathering these terms one obtains equation for balance of the total number of clusters $n$ 
\begin{equation}
N(1-\rho) - \sum_{i=1}^N \frac{\mu_i}{\nu} n_i i+n_{N-1}=2n_R.
\label{eq:baln}
\end{equation}
Again we emphasise that the above result is exact -- no correlations were neglected. 
Finite size of the system reflects in the appearance of $(2n_R-n_{N-1})$ instead of $2n$ in the respective formula in \cite{BiaCzAA}.

\subsection{Balance of $n_i$s}

{\bf Loss.} There are two modes. \\
(a) Enlarging - an empty cluster on the edge of an $i$-cluster becomes occupied. There are two such empty clusters except for the case when system contains a single cluster of length $N-1$. Hence, the respective rates are
\begin{eqnarray}
\sim & 2 \nu \frac{n_i}{N} \quad \quad \quad & i=1,\ldots,N-2, \\
\sim & \nu \frac{n_{N-1}}{N} \quad \quad & i=N-1. 
\end{eqnarray}
(b) Relaxation rate for any $i=1,\ldots,N$ is given by  
\begin{equation}
\sim \mu_i \frac{i n_i}{N}.
\end{equation} 

{\bf Gain.} Again, there are two modes.\\
(a) Enlarging. For $N\geq3$, there are following rates depending on the size $i$ of the cluster
\begin{eqnarray}
\sim & \nu(1-\rho) - 2\nu\frac{n_R}{N} + \nu \frac{n_1^0}{N}, \quad  & i=1, \\
\sim & 2\nu\frac{n_{i-1}}{N} \alpha^E_{i-1} \quad \quad \quad \quad  & 2\leq i \leq N-1, \\
\sim & \nu\frac{n_{N-1}}{N}, \quad \quad \quad \quad \quad \quad \quad \quad & i=N, \label{eq:gecN}
\end{eqnarray}
where $\alpha_E(i)$ is a probability that the size of empty cluster adjacent to the $i$-cluster is bigger than $1$. 
It is clear that 
\begin{equation}
\alpha^E_{N-2}=1 \quad \text{and} \quad \alpha^E_{N-1}=0.
\end{equation}
Formula \eqref{eq:gecN} does not have a factor $2$, because there is only one empty cluster (of size $1$).

(b) Merger of two clusters up to the cluster of size $i$. Two clusters: one of size $k\in \{1,2,\ldots,(i-2)\} $ and the other of size $((i-1)-k)$ will be combined if the ball fills an empty cell between them. 
$$ \hookrightarrow \cdots | \quad | \underbrace{ \overbrace{\bullet \ | \cdots | \ \bullet }^k | \quad | 
\overbrace{ \bullet \ | \bullet | \cdots | \ \bullet}^{(i-1-k)}  }_{i} | \quad | \cdots \hookleftarrow $$
The probability is proportional to the number of empty $1$-clusters between $k$-cluster and $(i-1-k)$-cluster,
\begin{equation}
\sim \nu\frac{n_{1}^0}{N} \gamma^E_{i}  \quad \quad \quad 3 \leq i \leq N-1,
\end{equation}
where $\gamma^E_{i}$ is a probability of such merger.  
For $i=N$ there is a single cluster in the lattice (there are no two clusters to merge) - 
filling the gap between ends of $(N-1)$-cluster is already considered in (a).

Gathering the terms, one obtains
\begin{eqnarray}
n_1 &=&\frac{1}{\frac{\mu_1}{\nu}+2} \left( N(1-\rho) -2n_R + n_1^0 \right), \label{eq:baln1} \\
n_2 &=& \frac{1}{2\frac{\mu_2}{\nu}+2} 2n_1 \alpha^E_{1},  \label{eq:baln2}\\
n_i &=&   \frac{1}{\frac{\mu_i}{\nu}i+2} \left( 2n_{i-1} \alpha^E_{i-1} + n_1^0 \gamma^E_{i} \right), \label{eq:balni}\\
n_{N-1} &=&   \frac{1}{\frac{\mu_{N-1}}{\nu}(N-1)+1} \left( 2n_{N-2} + n_1^0 \gamma^E_{N-1} \right), \label{eq:balnN-1}\\ 
n_N &=&  \frac{1}{\frac{\mu_N}{\nu}N} n_{N-1},   \label{eq:balnN}
\end{eqnarray} 
where $ \quad 3\leq i\leq(N-2)$.

The last equation \eqref{eq:balnN} has simple explanation. The state with all cells being occupied (corresponding to $n_N$) 
can be achieved only from the state with a single empty cell (corresponding to $n_{N-1}$) with probability $\nu (1/N) $.
On the other hand, the automaton leaves the state with all cells being occupied with probability $\mu_N$. 

Note that equations \eqref{eq:baln1} and \eqref{eq:balnN} are exact. 
Correlations in the systems reflect in appearing of multipliers $\alpha^E_{i}$ and $\gamma^E_{i}$. Their values 
depends on possible configurations of states of the automaton. 
As shown in the Appendix, for $N\geq 5$ exact formulas for $\alpha^E_{i}$ and $\gamma^E_{i}$ as functions of $n_i$s do not exist.
Hence, it is necessary to propose approximated formulas. 
 
A mean field type approximation for $\alpha^E_{i}$ is 
\begin{equation}
\alpha^E_{i} \approx  \alpha^A_{i} = \left( 1- \frac{n_1^0}{\sum_{k=1}^{N-i}n_k^0} \right).
\label{eq:alphaEapprox}
\end{equation}
For a given cluster of size $i$, the probability of appearance of an empty cluster of size $1$ is calculated as proportional
to the number of empty  $1$-clusters divided by the sum of the numbers of all empty clusters with size not exceeding $N-1$, because there is no room for larger.

When merger of two clusters up to a cluster of size $i$ is considered, the room denoted by $A$ is of size $(N-2-(i-1-k))$ and the room denoted by $B$ is of size $(N-2-k)$ - see a diagram below.
$$ \hookrightarrow \overbrace{ \underbrace{\cdots | }_{B}  \quad |  \underbrace{\bullet \ | \cdots | \ \bullet }_k}^A | \quad | 
\underbrace{ \overbrace{ \bullet \ | \bullet | \cdots | \ \bullet}^{(i-1-k)}  | \quad  \overbrace{| \cdots }^A }_B \hookleftarrow$$
Hence a mean field type approximation for $\gamma^E_{i}$ is of the form
\begin{equation}
\gamma^E_{i} \approx \gamma^A_{i} = \sum_{k=1}^{i-2} \left( \frac{n_k} {\sum_{j=1}^{N-(i-1-k+2)}n_j}  
\cdot \frac{n_{i-1-k}}{\sum_{j=1}^{N-(k+2)}n_j} \right).
\label{eq:gammaEapprox}
\end{equation}
It is also instructive to consider another approximation 
\begin{equation}
\gamma^E_{i} \approx \gamma^{AR}_{i} = \sum_{k=1}^{i-2} \left( \frac{n_k}{n_R} \cdot \frac{n_{i-1-k}}{n_R} \right).
\end{equation}
Section \ref{sec:examples} contains quantitative estimation of proposed approximations.
Comparison of this approximation with exact results for small sizes $N$ is discussed in Section \ref{sec:conclusions}.

\subsection{Thermodynamic limit}

In the paper \cite{BiaCzAA} an assumption of independence of clusters was considered. 
To have it adequate, it is required that there are no limitations in space, like those encountered when
formulas \eqref{eq:alphaEapprox} and \eqref{eq:gammaEapprox} were considered. 
For systems that are big enough, i.e., when $N \longrightarrow \infty$, an empty cluster adjacent to a given $i$-cluster
can be of any size, and thus 
\begin{equation}
\alpha^E_{i} \approx  \alpha = \left( 1- \frac{n_1^0}{\sum_{k=1}^{\infty}n_k^0} \right) = \left( 1- \frac{n_1^0}{n} \right).
\end{equation}
This is consistent with the requirement that $n_i \longrightarrow 0$ when $i \longrightarrow \infty$, which is required 
to have moments of the $n_i$s convergent.
Similarly,
\begin{equation}
\gamma^E_{i} \approx \gamma(i) = \sum_{k=1}^{i-2} \left( \frac{n_k}{n} \cdot \frac{n_{i-1-k}}{n} \right).
\end{equation}
These formulas substituted into \eqref{eq:baln1}-\eqref{eq:balni} give the respective set of equations considered in  \cite{BiaCzAA}.
The same reasoning can be applied to balance equations.
The form of equation \eqref{eq:rho_balance} is left unchanged under the limit. For equation \eqref{eq:baln},
$(2n_R-n_{N-1}) \longrightarrow 2n$, and it becomes of the form presented in \cite{BiaCzAA}.

\section{Special cases} \label{sec:cases}

For fixed form of rebound parameters equations describing the automaton can be written in more specific form.
This is the case for balance equations \eqref{eq:rho_balance} and   \eqref{eq:baln}, as well as for  formulas for average cluster size 
\begin{equation}
\left\langle i \right\rangle = \frac{\sum_{i=1}^N n_i i }{\sum_{i=1}^N n_i} = \frac{N\rho}{n_R+n_N}
\label{eq:avcluster}
\end{equation}
and average avalanche size
\begin{equation}
\left\langle w \right\rangle = \frac{\sum_{i=1}^N \mu_i n_i i^2 }{\sum_{i=1}^N \mu_i n_i i}.
\label{eq:avavlanche}
\end{equation}
 We emphasize, these formulas are exact -- correlations are encountered.
We consider three special cases investigated in detail and illustrated by examples below.

\subsection{$\mu=const.$}

For $\mu=const.$ and $\nu=const.$ equation \eqref{eq:rho_balance} is of the form
\begin{equation}
(1-\rho)= \frac{1}{N} \frac{\mu}{\nu}\sum_{i=1}^N  n_i i^2.
\label{eq:rho_balanceA}
\end{equation} 
and equation \eqref{eq:baln}
\begin{equation}
N(1-\rho(1+\frac{\mu}{\nu}))+n_{N-1}=2n_R.
\label{eq:balnA}
\end{equation}
Also formulas for $\left\langle i \right\rangle$ and $\left\langle w \right\rangle$ are simplified only a little. 

\subsection{$\mu(i)=\delta/i$ where $\theta=\delta/\nu=const.$}

Equation \eqref{eq:rho_balance} is of the form
\begin{equation}
(1-\rho)= \theta \rho,
\label{eq:rho_balanceB}
\end{equation} 
hence the density is given by remarkably neat (end exact) formula
\begin{equation}
\rho=\frac{1}{1+\theta}.
\label{eq:densityB}
\end{equation}
Note that there is no dependence on the size of the system $N$; for $N \longrightarrow \infty$ it remains the same.

Equation \eqref{eq:baln} can be written as
\begin{equation}
N\frac{\theta}{1+\theta} = (2+\theta)n_R,
\label{eq:balnB}
\end{equation}
where we use equations \eqref{eq:balnN}  and \eqref{eq:densityB}.
Hence the formula for $n_R$ is of the form
\begin{equation}
n_R=N\frac{\theta}{(\theta+1)(\theta+2)}
\label{eq:nRB}
\end{equation}
in direct analogy with $n$ in  $N \longrightarrow \infty$ case \cite{BiaCzAA}. 
Thus, $n_R$ plays the role of $n$, as indicated also in balance of $n_1$ equation \eqref{eq:baln1}. 
The formula for $n$ is 
\begin{equation}
n=n_R+n_N= N\frac{\theta (1+\varepsilon)}{(\theta+1)(\theta+2)} \quad \text{where} \quad \varepsilon=\frac{n_N}{n_R}.
\label{eq:nB}
\end{equation}

The average cluster size is given by 
\begin{equation}
\left\langle i \right\rangle = \frac{1}{1+\varepsilon}\left( 1+\frac{2}{\theta} \right). 
\label{eq:avclusterB}
\end{equation}
The average avalanche size is equal to the average cluster size 
\begin{equation}
\left\langle w \right\rangle = \left\langle i \right\rangle,
\label{eq:avavlancheB}
\end{equation}
because each cluster has the same probability to be removed from the lattice.

The above formulas are exact (include correlations) and have good thermodynamic limit ($\varepsilon \longrightarrow 0$).
Note also that variables $\rho$ and $n_R$ depend on single parameter $\theta$. Formulas with dependence on $\theta$ can be rewritten as functions of density $\rho$. 

\subsection{$\mu(i)=\eta/{i^2}$ and $\chi=\sigma/\nu=const.$}

Equation \eqref{eq:rho_balance} is of the form
\begin{equation}
N(1-\rho)= \chi (n_R+n_N).
\label{eq:rho_balanceC}
\end{equation} 

Equation \eqref{eq:baln} can be  written as
\begin{equation}
N(1-\rho)=2n_R-\chi\frac{1}{N}n_N+\chi\sum_{i=1}^N \frac{n_i}{i}
\label{eq:balnC}
\end{equation}
where equation \eqref{eq:balnN} is used, namely  $n_{N-1}=\chi\frac{1}{N}n_N$.

The average cluster size 
\begin{equation}
\left\langle i \right\rangle = \chi \frac{\rho}{1-\rho},
\label{eq:avclusterC}
\end{equation}
and the average avalanche size
\begin{equation}
\left\langle w \right\rangle = \frac{1}{\left( 1+ \frac{\sigma}{N} \right) - (1-\sigma)\frac{2}{\chi}}, 
\label{eq:avavlancheC}
\end{equation}
where
\begin{equation}
\sigma=\frac{n_N}{n_R+n_N}=\frac{\varepsilon}{\varepsilon+1}.
\end{equation}
Note that also these formulas are exact.

\section{Finite RDA as a Markov chain} \label{sec:Markov}

\subsection{General settings}

\begin{table}[t]%
\begin{tabular}{c|c|c|c}
 state number &  example\footnote{Other states differ by translations.} & multiplicity & contrib. to   \\
\hline\hline
1& $ \hookrightarrow | \ \ \ | \ \ \ | \ \ \ | \hookleftarrow $	& 1 & $ n_3^0 $ \\
\hline
2 & $ \hookrightarrow | \ \ \ | \ \ \ | \bullet | \hookleftarrow $	& 3 & $ n_1, n_2^0 $  \\
\hline
3 & $ \hookrightarrow | \ \ \ | \bullet | \bullet | \hookleftarrow $	& 3 & $ n_2, n_1^0 $ \\
\hline
4 & $ \hookrightarrow | \bullet | \bullet | \bullet | \hookleftarrow $	& 1 & $ n_3 $ 
\end{tabular}
\caption{States  for the size of the lattice $N=3$.}
\label{tab:N3}
\end{table}

Finite Random Domino Automaton is a Markov chain, hence we use standard knowledge to solve
several examples for small $N$ and derive a number of formulas for time aspects of the evolution of the system.

In general, for the lattice of size $N$ there are $2^N$ states, because each of $N$ cells may be empty or occupied. 
For $N=4$, an exemplary state is  
\begin{displaymath}
\hookrightarrow | \ \ \ | \bullet | \ \ \ | \bullet | \hookleftarrow	
\end{displaymath}
where assumed periodic boundary conditions are depicted by hook-arrows.

For periodic boundary conditions it is irrelevant to distinguish between 
states which differ by a translation only. Hence, in example, we consider the following states equivalent:
\begin{displaymath}
	\hookrightarrow | \ \ \ | \bullet | \ \ \ | \bullet | \bullet | \hookleftarrow 
\quad \quad \equiv \quad \quad 
\hookrightarrow | \bullet | \ \ \ | \bullet | \ \ \ | \bullet | \hookleftarrow
\end{displaymath}
Thus states $a_i$ are defined up to translational equivalence (see Tables \ref{tab:N3} and \ref{tab:N5}). 
The label numbers are assigned to the states, as shown in tables - no exact rule is applied.

\begin{table}[b]%
\begin{tabular}{c|c|c|c}
 state number &  example\footnote{Other states differ by translations.} & multiplicity & contrib. to   \\
\hline\hline
1 & $ \hookrightarrow | \ \ \ | \ \ \ | \ \ \ | \ \ \ | \ \ \ | \hookleftarrow $	& 1 & $ n_5^0 $ \\
\hline
2 & $ \hookrightarrow | \ \ \ | \ \ \ | \ \ \ | \ \ \ | \bullet | \hookleftarrow $	& 5 & $ n_1, n_4^0 $ \\
\hline
3 & $ \hookrightarrow | \ \ \ | \ \ \ | \ \ \ | \bullet | \bullet | \hookleftarrow $	& 5 & $ n_2, n_3^0 $ \\
\hline
4 & $ \hookrightarrow | \ \ \ | \ \ \ | \bullet | \ \ \ | \bullet | \hookleftarrow $	& 5 & $ n_1, n_1^0, n_2^0 $ \\
\hline
5 & $ \hookrightarrow | \ \ \ | \ \ \ | \bullet | \bullet | \bullet | \hookleftarrow $	& 5 & $ n_2, n_2^0 $ \\
\hline
6& $ \hookrightarrow | \ \ \ | \bullet | \ \ \ | \bullet | \bullet | \hookleftarrow $	& 5 & $ n_1, n_2, n_1^0 $ \\ 
\hline
7 & $ \hookrightarrow | \ \ \ | \bullet | \bullet | \bullet | \bullet | \hookleftarrow $& 5 & $ n_4, n_1^0 $ \\
\hline
8 & $ \hookrightarrow | \bullet | \bullet | \bullet | \bullet | \bullet | \hookleftarrow $ & 1 & $ n_5 $

\end{tabular}
\caption{States for the size of the lattice $N=5$.}
\label{tab:N5}
\end{table}

Further reduction of the number of states using reflections can be done, but it is not very efficient procedure.
We do not perform it, keeping symmetrical states separate. They deliver a simple computation check - their probabilities are necessarily equal. 

Such space of states for the finite random domino automaton is irreducible, aperiodic and recurrent.
Transition matrix $\mathbf{P}$ is defined by
\begin{equation}
[ \mathbf{P} ]_{ij} = \text{probability of transition} \quad a_i \longrightarrow a_j
\label{eq:transMat}
\end{equation}

For $N=3$ the transition matrix is of the form
\begin{equation}
\mathbf{P} =
\left( \begin{array}{cccc}
1-\nu & \nu & 0 & 0 \\
\frac{\mu_1}{3} & 1- \frac{\mu_1}{3}-\frac{2\nu}{3} & \frac{2\nu}{3} & 0 \\
\frac{2\mu_2}{3} & 0 &  1- \frac{2\mu_2}{3}-\frac{\nu}{3} &  \frac{\nu}{3} \\
 \mu_3   & 0 & 0 & 1-\mu_3
\end{array} \right)
\label{eq:transM3}
\end{equation}
where entries are found from analysis of transition probability of all possible states $a_i$
(see Tab.\ref{tab:N3}).

For $N=5$ the transition matrix is 
\begin{widetext}
\begin{equation}
\mathbf{P} = \frac{1}{5}
\left( \begin{array}{cccccccc}
 5-5\nu &   5\nu&       0&           0&               0&            0&               0&          0 \\ 
 \mu_1  &   5-\mu_1-4\nu & 2\nu       &   2\nu&            0&            0&               0&          0\\
 2\mu_2 &    0&          5-2\mu_2-3\nu& 0&               2\nu&         \nu&              0&          0\\
   0    &        2\mu_1&      0&            5-2\mu_1-3\nu&  \nu&            2\nu&            0&          0\\
  3\mu_3&    0&          0&            0&               5-3\mu_3-2\nu& 0&               2\nu&       0\\
     0  &        2\mu_2&      \mu_1&          0&               0&            5-2\mu_2-\mu_1-2\nu& 2\nu&        0\\
  4\mu_4&    0&          0&            0&               0&            0&               5-4\mu_4-\nu& \nu\\
  5\mu_5&    0&          0&            0&               0&            0&               0&          5-5\mu_5 
\end{array} \right)
\label{eq:transM5}
\end{equation}
\end{widetext}

Stationary distribution is given by
\begin{equation}
v \cdot P=v.
\label{eq:MP}
\end{equation}

The number of states increase rapidly with $N$:
for $N=6$ there are $14$ states, 
for $N=7$ there are $20$ states and for $N=10$ there are $108$ states.
The number of states for any $N$ is bigger than $2^N/N$, because translational symmetry of states is at most $N$, but 
always there are states with smaller symmetry, like empty state and fully occupied state.   
Thus practical usage of Markov chain settings for calculations is rather limited. This is one of the reasons for 
developing more "handy" framework, like presented in \cite{BiaCzAA} and here.
On the other hand, Markov chains can be used for illustrations and justifications of some properties, as presented below.

\subsection{Expected time of return}

As system evolves, it hits a given state many times. Here we consider expected value of the time of return from  
state with density $\rho=0$  to itself and next from the state with $\rho=1$ to itself.
 
Starting from  state $1$ (state with $\rho=0$) the next state (different from state $1$) contains a single $1$-cluster only.
This state - denoted by label $2$ - has density $\rho=1/N$. Expected time for this change is $1/\nu$.

Let $\tau_i$ be the expected time to hit state $1$ starting in state $i$. Then $\tau_1=0$ and for $i\neq 1$
\begin{eqnarray}
\tau_i &=&\mathbb{E}( \text{time to hit} \ 1 \ | \ \text{start in} \ i  )  \nonumber \\
&=& 1 + \sum_{k} p_{ik} \mathbb{E}(1 | k) = 1 + \sum_k p_{ik}\tau_k,
\label{eq:rettime}
\end{eqnarray}
where   $\mathbb{E}(1|k) =\mathbb{E}( \text{time to hit} \ 1 \ | \ \text{start in} \ k  )$.
After solving this system of equations, the return time is
\begin{equation}
t_{1\rightarrow 1}= 1/\nu + \tau_2.
\label{eq:rettime11}
\end{equation}

Similarly, for state with $\rho=1$ (state $L$) the next state (different from state $L$) is the empty state (with $\rho=0$) 
and 
\begin{equation}
t_{L \rightarrow L}= 1/\mu_N + \hat\tau_1,
\label{eq:rettimeL}
\end{equation}
where $\hat\tau_1$ is the expected time to hit state $L$ starting in state $1$.
The respective equation to determine $\hat\tau_i$ for $i\neq L$ reads
\begin{equation}
\hat\tau_i  = 1 + \sum_k p_{ik} \hat\tau_k,
\label{eq:hatrettime}
\end{equation}
and obviously $\hat\tau_L=0$.

Note that the expected time $t_{L \rightarrow L}$ is equal to expected time of return from state $1$ to state $1$ through state $L$:
\begin{equation}
t_{L \rightarrow L}=t_{1\rightarrow L \rightarrow 1}.
\label{eq:timell1l1}
\end{equation} 

\begin{table}[b]%
\begin{tabular}{l|c}
   probability of  & value \\
\hline
rebound -- occupied cell & $\frac{1}{N}\sum_{i=1}^N n_i i (1-\mu_i)$\\
rebound -- empty cell & $(1-\rho)(1-\nu) $ \\
occupation of empty cell &  $(1-\rho)\nu$\\
trigerring an avalanche &  $\frac{1}{N}\sum_{i=1}^N \mu_i n_i i$  
\end{tabular}
\caption{Probabilities of all four possibilities occurring in a single time step during evolution of the automaton.}
\label{tab:probrebocc}
\end{table}

The expected time between two consecutive avalanches is  
\begin{equation}
t_{av}= \frac{\left\langle w \right\rangle + 1}{1-P_r}, 
\label{eq:timeavalanche}
\end{equation}
where $P_r$ is the probability that the incoming ball is rebounded both form empty or occupied cell:
\begin{equation}
P_r = (1-\rho)(1-\nu) + \frac{1}{N}\sum_{i=1}^N n_i i (1-\mu_i).
\label{eq:Pr}
\end{equation}
Note that $(1-P_r)$ is equal to the sum of probability of triggering an avalanche and probability that an empty cell becomes occupied, hence
\begin{equation}
P_r + \frac{1}{N}\sum_{i=1}^N \mu_i n_i i +  (1-\rho)\nu=1.
\label{eq:Pz}
\end{equation}
Formula \eqref{eq:timeavalanche} can be derived as follows. In time between two consecutive avalanches,
on average, $(t_{av} (1-P_r)-1)$ cells become occupied in the system -- it receives one ball per a time step,  part of them are 
rebounded and one ball triggers the avalanche. 
An avalanche is reducing the number of occupied cells by $\left\langle w \right\rangle$. These two quantities compensate each other, giving \eqref{eq:timeavalanche}.

On the other hand, the expected time between two consecutive avalanches is equal to the inverse of the probability of triggering an avalanche
\begin{equation}
t_{av} =  \left( \frac{1}{N}\sum_{i=1}^N \mu_i n_i i \right)^{-1}.
\label{eq:tavav}
\end{equation}
Both expressions given in   \eqref{eq:timeavalanche} and \eqref{eq:tavav} are equal to each other.

\subsection{Frequency distribution of avalanches}

The probability of states obtained from condition \eqref{eq:MP} allows to determine the distribution of frequency of avalanches.
The frequency $f_i$ of the avalanche of size $i$ is given by the sum of products of probabilities $v_k$ of state $k$ and respective transition probability $p_{kj}$ to the appropriate states $j$ for all states that transition $k \longrightarrow j$ 
produce the avalanche of size $i$. 
 
For example, for $N=5$, as can be seen in Table \ref{tab:N5}, transitions $2 \longrightarrow 1$, $4 \longrightarrow 2$ and $6 \longrightarrow 3$ result in an avalanche of size $1$, 
transitions $3 \longrightarrow 1$ and $6 \longrightarrow 2$ give an avalanche of size $2$, transition $5 \longrightarrow 1$ of size $3$,
$7 \longrightarrow 1$ of size $4$ and $8 \longrightarrow 1$ of size $5$.
Hence 
\begin{eqnarray}
f_1&=& v_2 \mu_1/5+v_4 2\mu_1/5+ v_6 \mu_1/5, \\
f_2&=& v_3 2 \mu_2/5+v_6 2 \mu_2/5, \\ 
f_3&=& v_5 3 \mu_3/5,  \\
f_4&=& v_7 4 \mu_4/5, \\
f_5&=& v_8 \mu_5, 
\label{eq:avalanchefreq}
\end{eqnarray} 
where respective $p_{kj}$ are taken from transition matrix \eqref{eq:transM5}.

The average time $t_i$ between two avalanches of size $i$ is given by 
\begin{equation}
t_i = 1/f_i,
\label{eq:avallanchetime}
\end{equation}
in particular, for a maximum size $N$ 
\begin{equation}
t_{L \rightarrow L}=t_N.
\label{eq:tlltn}
\end{equation}

The average time between (any) consecutive avalanches given by formula \eqref{eq:tavav} may be also calculated as
\begin{equation}
t_{av}= \left( \sum_{i=1}^N  t_i^{-1}\right)^{-1},
\label{eq:tavsumti}
\end{equation}
because the probability of avalanche of any size is just a sum of probabilities of all possible avalanches. 
In this way one can calculate also average time between any two consecutive avalanches of prescribed size - for example, size $4$ and $5$ (or any other subset of possible sizes). 

\section{Examples} \label{sec:examples}

Below we present several examples to illustrate properties of finite RDA as well as to demonstrate application of 
the schemes outlined above.

\subsection{$N=3$}

This is the simplest non-trivial, worm-up example.
For $N=3$ the general results -- i.e., for arbitrary $\mu_1$, $\mu_2$, $\mu_3$ and $\nu$ -- can be calculated explicitly. Usage of equations \eqref{eq:baln1}-\eqref{eq:balnN} leads to exact results as presented below (see Appendix). The same can be also
obtained from Markov chains framework.
Equations \eqref{eq:rho_balance},   \eqref{eq:baln}, and \eqref{eq:balnN} 
give
\begin{eqnarray}
n_1  &=&  {3\left(\frac{\mu_2}{\nu} + \frac{1}{2}\right)}/D, \\
n_2  &=&  {3}/D,   \\
n_3  &=&  \left({\frac{\nu}{\mu_3}}\right)/D,
\label{eq:exactsolN3}
\end{eqnarray}
where 
\begin{equation*}
D = \frac{11}{2} + \frac{\mu_1}{2\nu} +  5\frac{\mu_2}{\nu} + \frac{\mu_1 \mu_2}{\nu^2} + \frac{\nu}{\mu_3}.
\end{equation*}
From inspecting of Table \ref{tab:N3} it is evident that $n_1^0=n_2$, $n_2^0=n_1$ and $ n_3^0 = 1- n_1 -n_2 -n_3$
(all posibilities sum up to 1), hence 
\begin{equation}
n_3^0  =  \left({1 + \frac{\mu_1}{2\nu} +  2\frac{\mu_2}{\nu} + \frac{\mu_1\mu_2}{\nu^2}}\right)/D.
\end{equation}

General formulas for expected times of return are 
\begin{eqnarray}
t_{1 \rightarrow 1} &=& 
\frac{1}{\nu} \left( 1+ \frac{2\nu^2+9\mu_3\nu+6\mu_2\mu_3}{\mu_3(\mu_1+2\nu)(2\mu_2+\nu)}  \right), \label{eq:n3ret1} \\
t_{L \rightarrow L} &=& 
\frac{1}{\nu} \left( \frac{\nu}{\mu_3} + \frac{11}{2} +\frac{\mu_1}{2\nu} + 5\frac{\mu_2}{\nu}+  \frac{\mu_1 \mu_2}{\nu^2}  \right).
\label{eq:n3retL}
\end{eqnarray}
The ratio $t_{L \rightarrow L}/t_{1 \rightarrow 1}$ is 
\begin{equation}
t_{L \rightarrow L}/t_{1 \rightarrow 1}= \frac{1}{2} \left(\frac{\mu_1}{\nu}+ 2\right)\left( \frac{2\mu_2}{\nu} + 1\right).
\label{eq:n3ttratio}
\end{equation}
Note that it does not depend on $\mu_3$. 
If the probability of triggering an avalanche of size $1$ and $2$ is small comparing to the probability of 
occupation of an empty cell (i.e., $\mu_1/\nu \approx 0$ and $\mu_2/\nu \approx 0$) 
then $t_{L \rightarrow L} \approx t_{1 \rightarrow 1}$. 
The next stage after the lattice is fully occupied is the empty state; hence, if these two average waiting times are comparable,
then they occur with comparable frequency. 
That means quasi-periodic like behaviour of the system: within average time $11/2\nu$ the lattice become fully occupied, 
then the triggering of an avalanche of maximal size $N$ occurs with average waiting time $1/\mu_3$. The same can be observed for bigger sizes $N$.

\begin{figure}[t]  
	\centering
	\includegraphics[width=4cm]{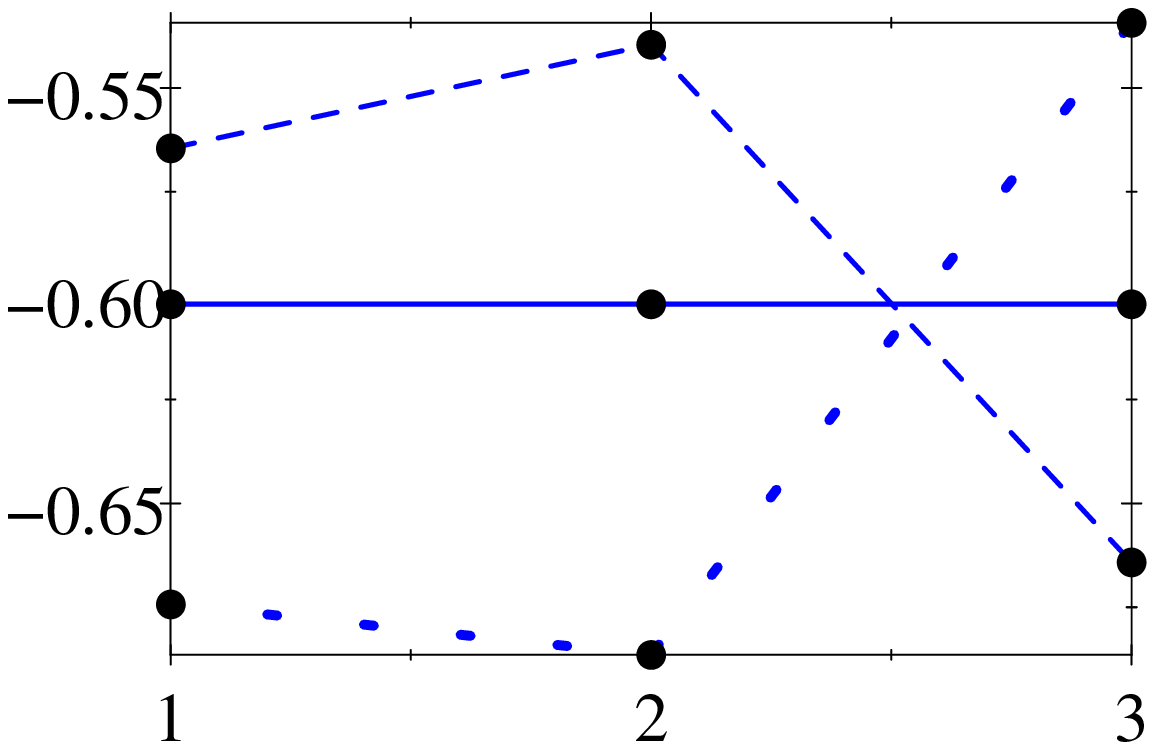} \quad 
	\includegraphics[width=4cm]{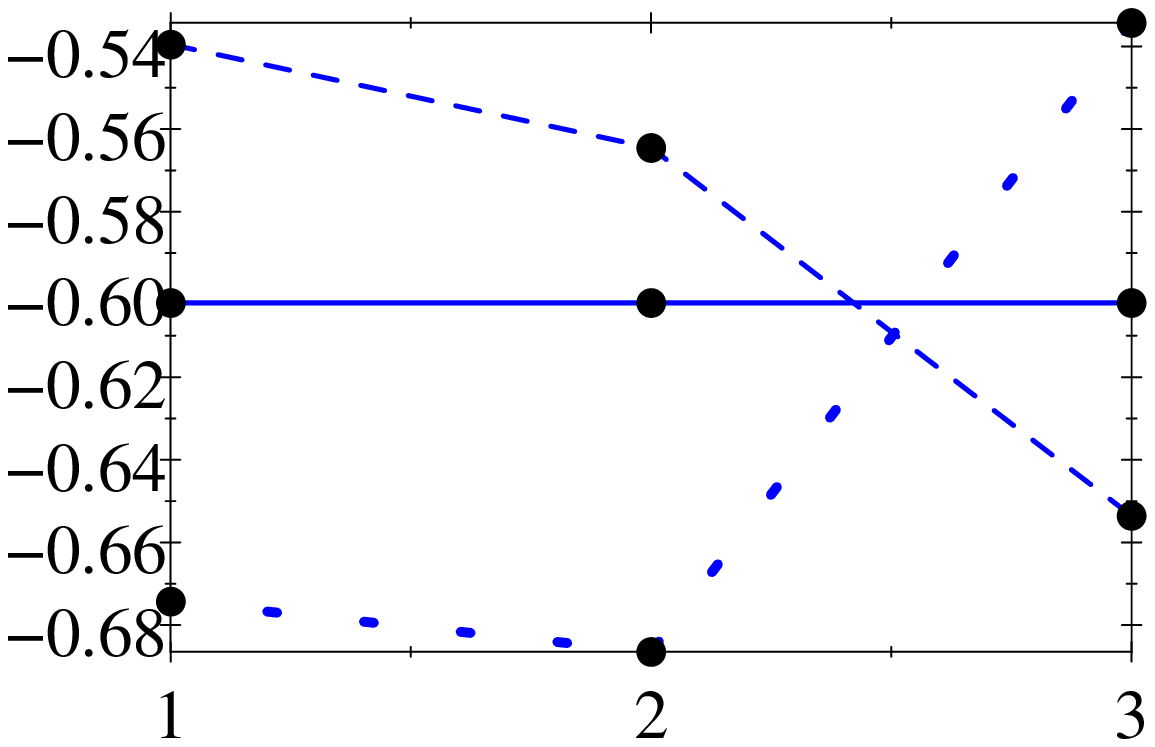} 
	\caption{Plot of the $Log_{10}$ of $n_i$s (left) and $n_i^0$s (right) vs. $i$ for $N=3$ in three cases:  ${\mu_i}=const.$  
	(dashed line), $\mu_i=\delta/i$ (solid line) and  $\mu_i=\sigma/i^2$ (dotted line). Rebound parameters are
	chosen to have density $\rho=1/2$ in all cases (see main text for respective values).} \label{fig:Fig1}
\end{figure}

\begin{table}[t]%
\begin{tabular}{c|c|c|c}
  & ${\mu_i}=const.$ &  $\mu_i=\delta/i$ & $\mu_i=\sigma/i^2$ \\
\hline
 $\left\langle i \right\rangle$ &    $1.9281668$    & $2$   &  $2.1134407$ \\
 $\left\langle w \right\rangle$ &    $2.2516538$    & $2$   &  $1.7226121$     
\end{tabular}
\caption{Average cluster size $\left\langle i \right\rangle$ and 
average avalanche size $\left\langle w \right\rangle$ for three different rebound parameters. Density $\rho=1/2$, 
the size of the lattice $N=3$.}
\label{tab:N3av}
\end{table}

Figure \ref{fig:Fig1} and Table \ref{tab:N3av} present examples of three types of dependence of rebound parameters on size $i$ of clusters considered in Section \ref{sec:cases}, 
each having the same density $\rho=1/2$ (with 8 digits accuracy).
To obtain this density we put for these three cases  $\mu/\nu=0.444118$ ($\mu=0.444118$, $\nu=1$), $\theta=1$ ($\delta=1$, $\nu=1$) and $\chi=2.113440690$ ($\eta=1$, $\nu=1/2.113440690$) respectively. As seen from Figure \ref{fig:Fig1} it is possible to obtain flat distribution for $\mu_i=\delta/i$ -- 
on that background, differences between the cases are clearly visible: $\mu_i=const.$ discriminate the existence of big clusters 
fostering big avalanches; the opposite is for $\mu_i=\sigma/i^2$. Average cluster size and avalanche size data presented in  Table \ref{tab:N3av} confirm this conclusion.

\begin{figure}[b]  
	\centering
	\includegraphics[width=4cm]{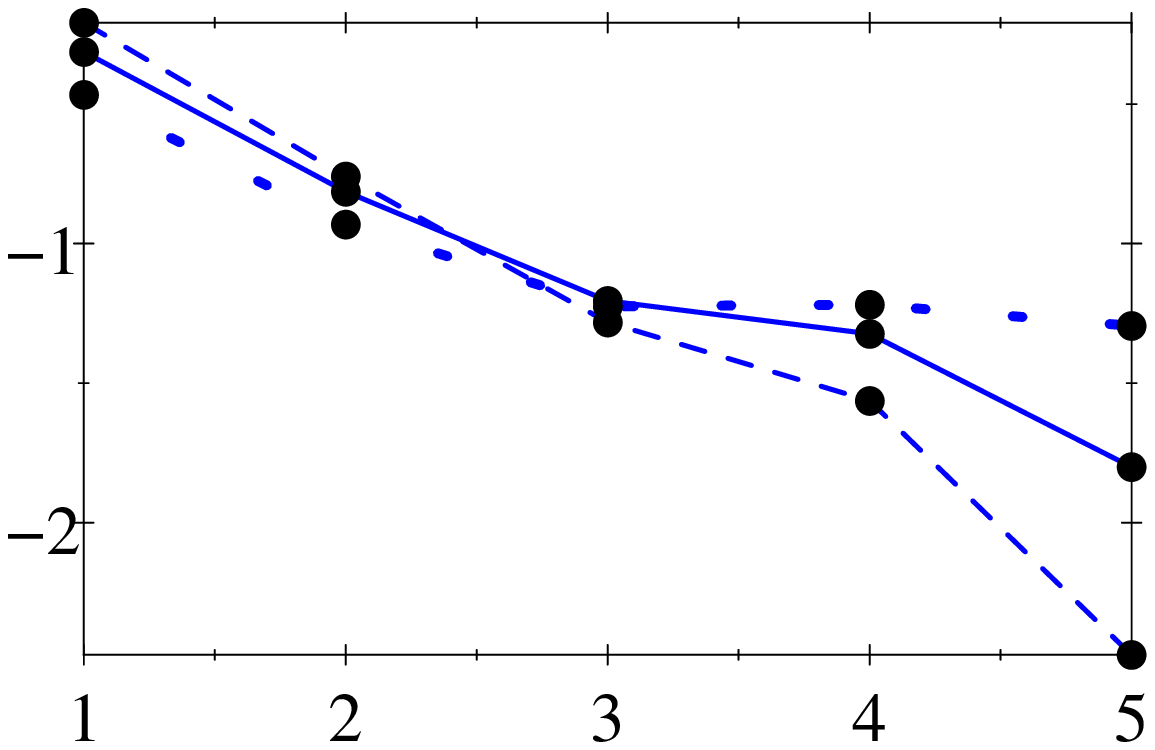} \quad 
	\includegraphics[width=4cm]{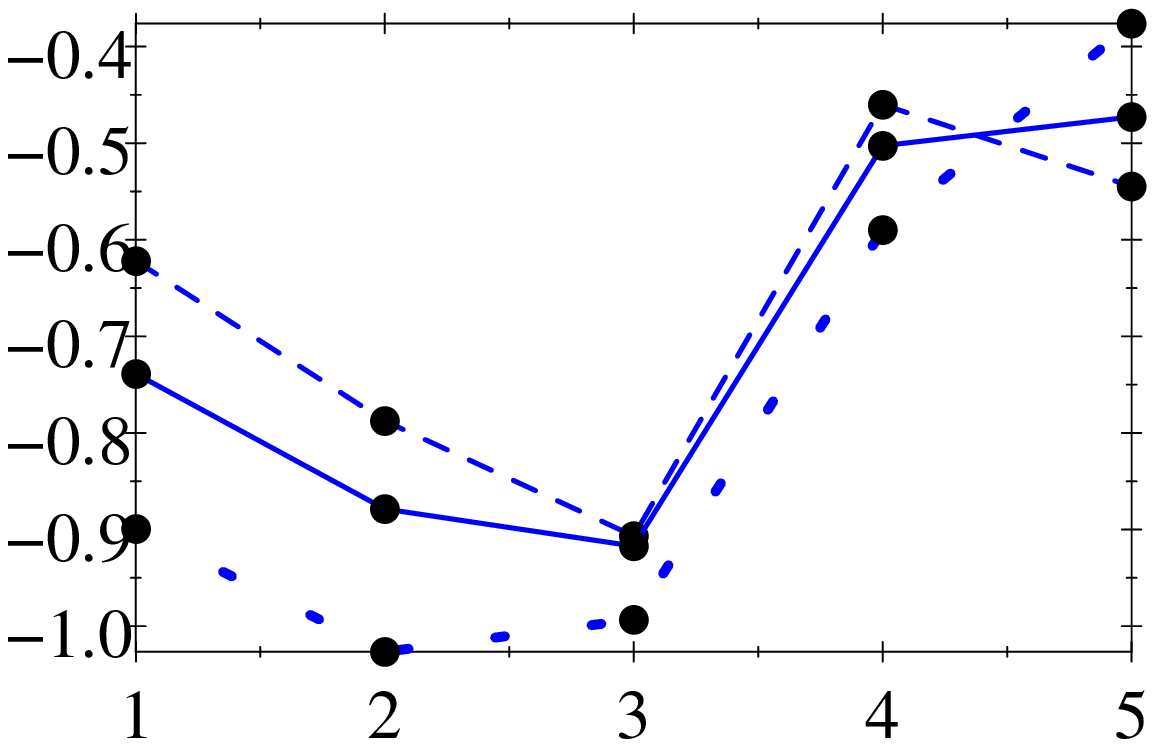} 
	\caption{Plot of the $Log_{10}$ of $n_i$s (left) and $n_i^0$s (right) versus $i$ for $N=5$ in three cases:  ${\mu_i}=const.$  
	(dashed line), $\mu_i=\delta/i$ (solid line) and  $\mu_i=\sigma/i^2$ (dotted line). Rebound parameters are
	chosen to have density $\rho=1/4$ in all cases (see main text for respective values).} \label{fig:Fig2a}
\end{figure}

\begin{table}[b!]%
\begin{tabular}{c|c|c|c}
  & ${\mu_i}=const.$ &  $\mu_i=\delta/i$ & $\mu_i=\sigma/i^2$ \\
\hline
 $\left\langle i \right\rangle$ &    $1.427017126$    & $1.632218845$   &  $1.985611461$ \\
 $\left\langle w \right\rangle$ &    $1.845355789$    & $1.632218845$   &  $1.41360643$       
\end{tabular}
\caption{Average cluster size $\left\langle i \right\rangle$ and 
average avalanche size $\left\langle w \right\rangle$ for three different rebound parameters. Density $\rho=1/4$, 
the size of the lattice $N=5$.}
\label{tab:N5av}
\end{table}

\begin{figure}[t]  
	\centering
	\includegraphics[width=4cm]{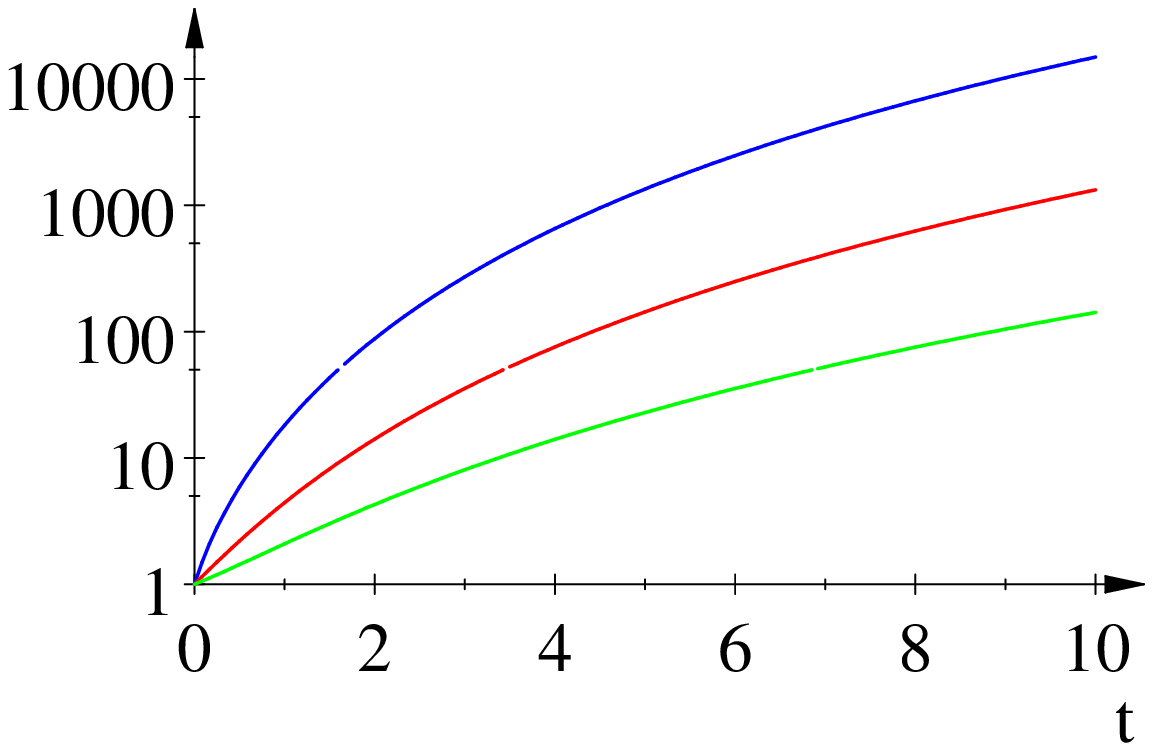} \quad 
	\includegraphics[width=4cm]{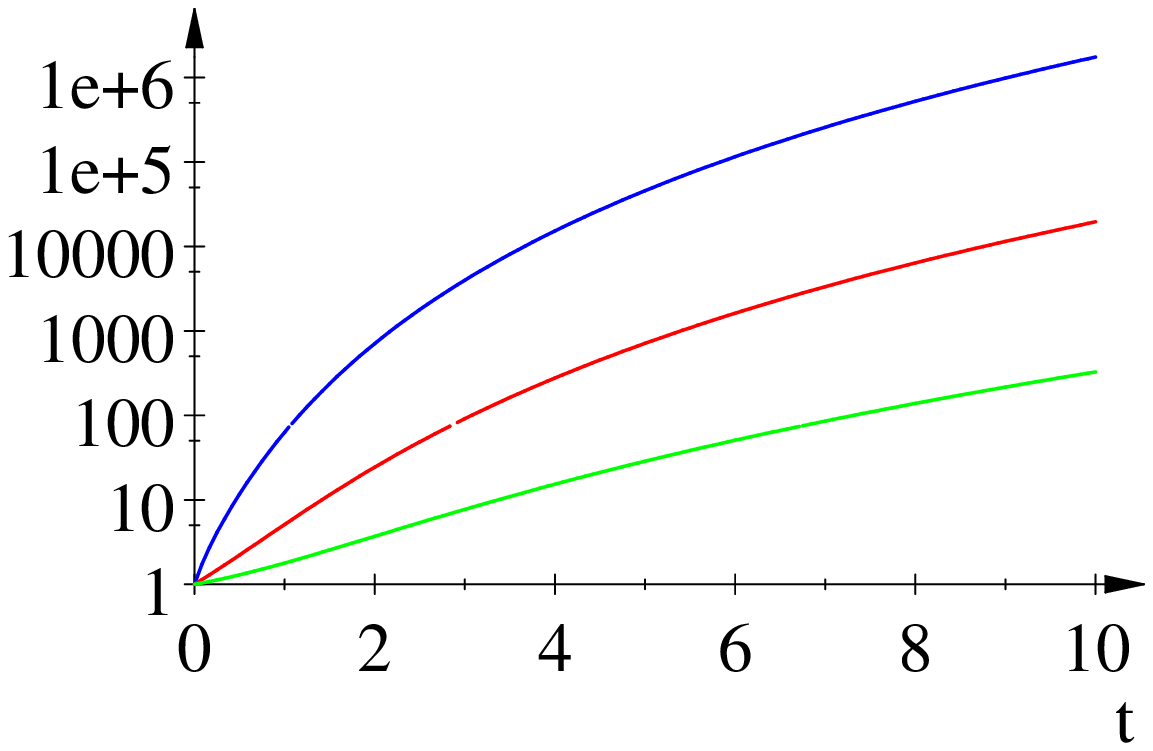}
	\caption{Ratio of return times ${t_{L \rightarrow L}}/{t_{1 \rightarrow 1}}$ for N=5 (left) and N=7 (right) for three cases: $\mu_i=const.$ (top),
	$\mu_i=\delta/i$ (middle) and  $\mu_i=\eta/i^2$ (bottom). Parameter $t$ is equal to $\mu/\nu$, $\delta/\nu$ and  $\eta/\nu$ respectively.} \label{fig:FigRet5}
\end{figure}

\begin{table}[b]%
\begin{tabular}{c|c|c|c}
  & ${\mu_i}=const.$ &  $\mu_i=\delta/i$ & $\mu_i=\sigma/i^2$ \\
\hline
 $R={t_{L \rightarrow L}}/{t_{1 \rightarrow 1}}$ &    $\approx52.212$    & $\approx35.441$   &  $\approx34.801$     
\end{tabular}
\caption{ Coefficient $R={t_{L \rightarrow L}}/{t_{1 \rightarrow 1}}$  for three different rebound parameters (see main text for details).
Density for all cases $\rho=1/4$, the size of the lattice $N=5$.}
\label{tab:N5tavR}
\end{table}

\subsection{$N=5$}

For $N=5$ it is impossible to write down  exact  equations \eqref{eq:baln1}-\eqref{eq:balnN} depending on values of  $n_i$s only -- see Appendix for details. The case can be solved as a Markov process, but obtained general formulas are relatively complicated. 
 
In this example we investigate properties of the system with density $\rho=1/4$.
Figure \ref{fig:Fig2a} and Table \ref{tab:N5av} compare results in three cases:
$\mu/\nu =16257/10000$ the  density $\rho=0.2500003184$; for $\theta=3$   the density  $\rho=0.25$ exactly;
and $\chi=5.95682$ gives the density $\rho=0.2500004527$.

General expressions for return times ${t_{1 \rightarrow 1}}$ and ${t_{L \rightarrow L}}$ as well as their ratio 
(presented in Appendix) are relatively complex. Note that the return times -- except of the dependence on $t$ -- are proportional to $1/\nu$. 
Below we specify the ratio ${t_{L \rightarrow L}}/{t_{1 \rightarrow 1}}$ in three cases:
for {$\mu_i=const$}, where $t=\mu/\nu$, it is equal to
\begin{equation}
\frac{24 t^6 + 154 t^5 + 413 t^4 + 586 t^3 + 467 t^2 + 182 t + 24}{24 t^2 + 54 t + 24},
\label{eq:n5ttratioc}
\end{equation}
for {$\mu_i=\delta/i$, where $\delta=const$} and $t=\delta/\nu$, it is equal to
\begin{equation}
\frac{4 t^6 + 40 t^5 + 169 t^4 + 395 t^3 + 550 t^2 + 432 t + 144}{56 t^2 + 192 t + 144},
\label{eq:n5ttratioci}
\end{equation}
and for {$\mu_i=\sigma/i^2$, where $\sigma=const$} and $t=\sigma/\nu$, is
\begin{equation}
\frac{2 t^6 + 39 t^5 + 304 t^4 + 1232 t^3 + 2840 t^2 + 3744 t + 2304}{496 t^2 + 2208 t + 2304}.
\label{eq:n5ttratiocii}
\end{equation}
In each case the ratio is a rational function of $t$, which is equal to $1$ for $t=0$ and asymptotically $\sim t^4$ for $t\longrightarrow \infty$. A generalisation of this observation is a Conjecture formulated in Section \ref{sec:conclusions}.
A comparison of these ratios is presented in left part of Figure \ref{fig:FigRet5}. 
Table \ref{tab:N5tavR} shows that for the cases discussed above  with average density $\rho=1/4$ the highest value of $R$ is for 
$\mu_i=const.$ and the smallest for $\mu_i=\sigma/i^2$ (not much different from the value for $\mu_i=\delta/i$).

\begin{figure}[t]  
	\centering
	\includegraphics[width=4cm]{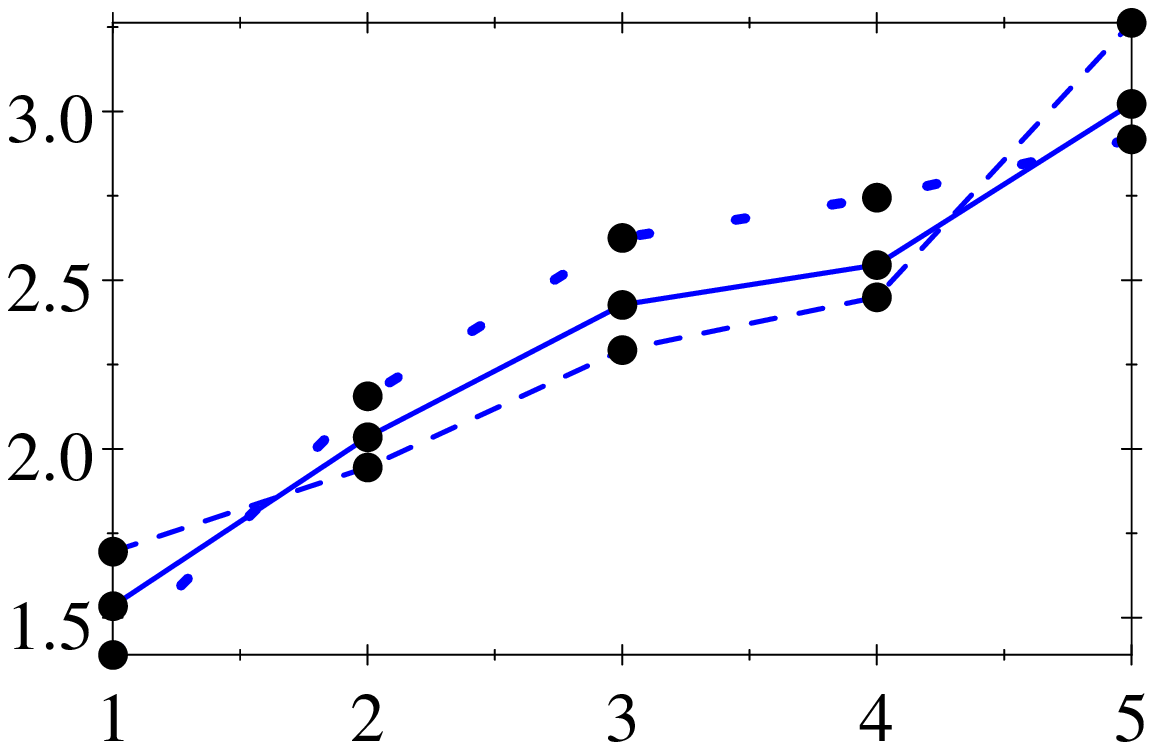} \quad 
	\includegraphics[width=4cm]{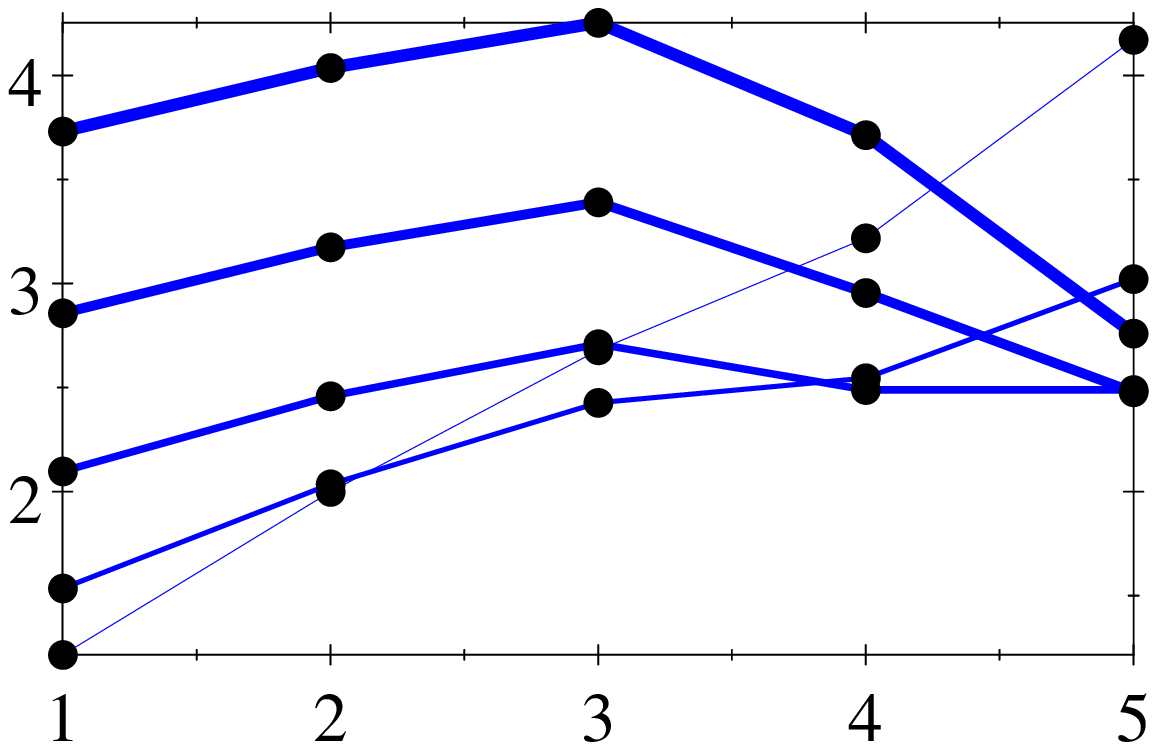} 
	\caption{
	{\bf Left.} Plot of $Log_{10}$ of $t_i$s vs. $i$ for three rebound parameters for fixed density $\rho=1/4$  
	 for $N=5$.	 Three cases:  ${\mu_i}=const.$  
	(dashed line), $\mu_i=\delta/i$ (solid line) and  $\mu_i=\sigma/i^2$ (dotted line). Rebound parameters are
	chosen to have density $\rho=1/4$ in all cases (see main text for respective values).
	{\bf Right.} Plot of $Log_{10}$ of $t_i$s versus $i$ for various densities for rebound parameter of the form $\mu_i=\delta/i$    
	 for $N=5$.	Densities  are chosen as $\frac{1}{10}, \frac{1}{4}, \frac{1}{2}, \frac{3}{4}, \frac{9}{10}$;
   thinner line corresponds to smaller density.
	} \label{fig:Figtimes}
\end{figure}

Average waiting times $t_i$ for avalanche of size $i$ can be also found. 
For example for $\mu_i=\delta/i$, where $\delta=const.$,  they are 
presented in the Appendix (equations \eqref{eq:wt1}-\eqref{eq:wt5}).
The average time between any two consecutive avalanches is
\begin{equation}
t_{av}=  \frac{4 t^5 + 48 t^4 + 237 t^3 + 603 t^2 + 762t + 360}{\nu t (4 t^4 + 36 t^3 + 121 t^2 + 168 t + 72)},
\label{eq:tav5ex}
\end{equation}
where $t=\delta/\nu$.
All these quantities are proportional to $1/\nu$. 
Figure \ref{fig:Figtimes} in the left panel presents waiting times $t_i$ in for fixed density $\rho=1/4$ in three cases mentioned above.  There are no big differences both in character of dependence of $t_i$ on $i$ and also values of $t_{av}$ do not differ much: 
for ${\mu_i}=const.$ average time is  $t_{av}\approx 24.60$,  for $\mu_i=\delta/i$ it  is   $\approx21.76$  and for 
 $\mu_i=\sigma/i^2$ it is  $\approx18.85$.
(Choosing parameters to have density $\rho=1/4$ we put $\nu=1/10$ for all cases.) 
 
Average waiting times $t_i, i=1,\ldots,5$ in the case $\mu_i=\delta/i$ for various densities are shown in the right panel of Figure \ref{fig:Figtimes}. For small densities the maximal waiting time $t_i$ is for $i=5$, while for bigger densities the maximum is for 
$i=3$.  Average waiting times range from $\approx13.57$ for $\rho={1}/{10}$ through
 $\approx21.76$, $\approx50.22$, $\approx145.01$ for densities ${1}/{4}$, $ {1}/{2}$, $ {3}/{4}$ respectively, up to $\approx441.60$  for density $\rho={9}/{10}$. (Again $\nu=1/10$ for all cases.)

\subsection{$N=7$}

\begin{figure}[t]  
	\centering
	\includegraphics[width=4cm]{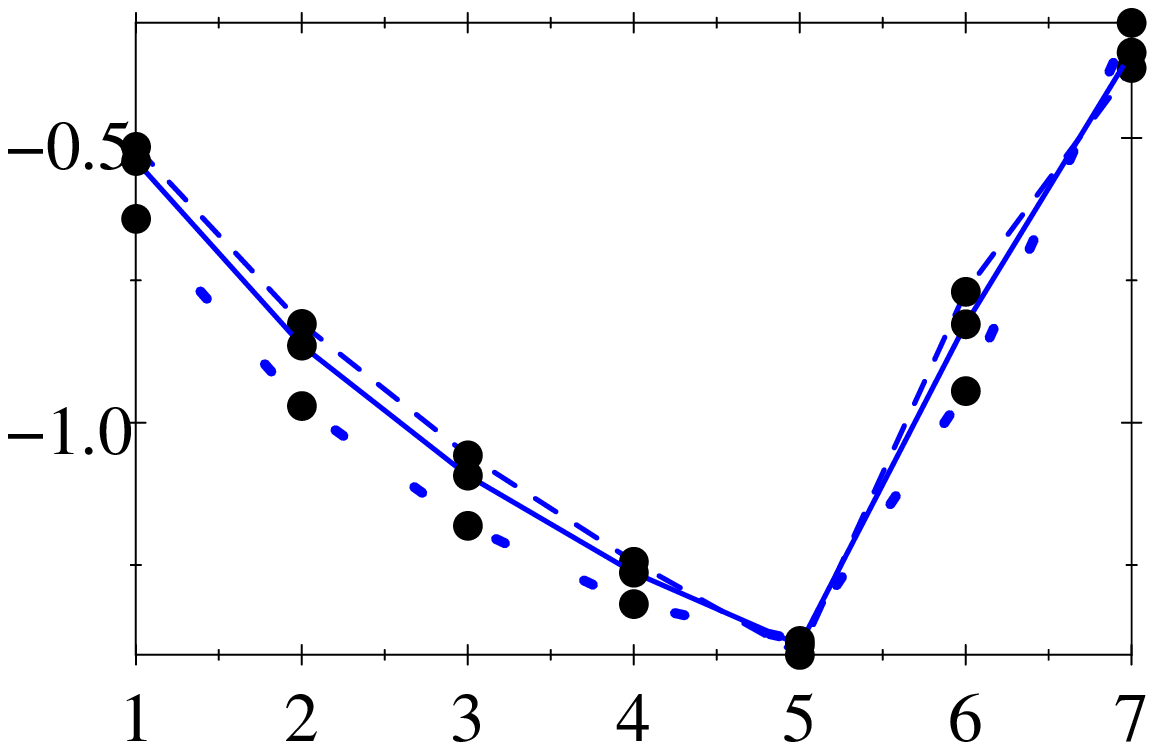} \quad 
	\includegraphics[width=4cm]{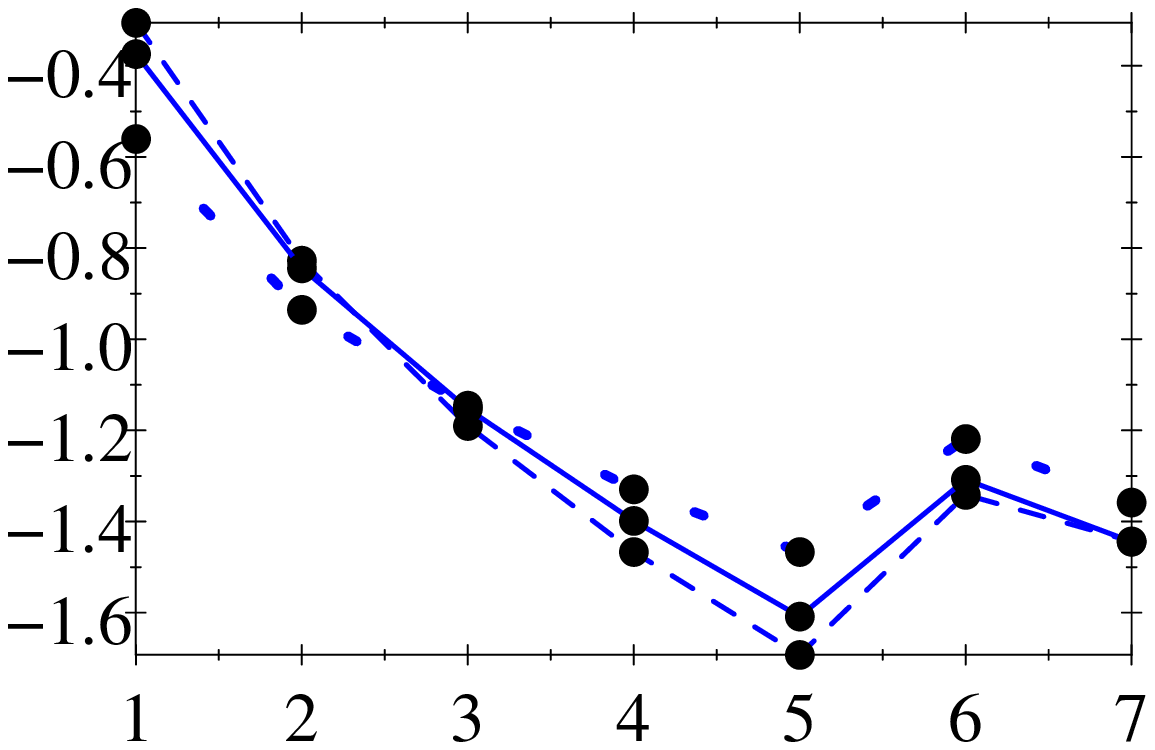} 
	\caption{Plot of the $Log_{10}$ of $n_i$s (left) and $n_i^0$s (right) versus $i$ for $N=7$ in three cases:  ${\mu_i}=const.$  
	(dashed line), $\mu_i=\delta/i$ (solid line) and  $\mu_i=\sigma/i^2$ (dotted line). Rebound parameters are
	chosen to have density $\rho=3/4$ in all cases (see main text for respective values).} \label{fig:Fig3a}
\end{figure}

\begin{table}[b]%
\begin{tabular}{c|c|c|c}
  & ${\mu_i}=const.$ &  $\mu_i=\delta/i$ & $\mu_i=\sigma/i^2$ \\
\hline
 $\left\langle i \right\rangle$ &    $4.274328495$    & $4.385371765$   &  $4.736665115$ \\
 $\left\langle w \right\rangle$ &    $5.767461682$    & $4.385371765$   &  $2.671314107$     
\end{tabular}
\caption{Average cluster size $\left\langle i \right\rangle$ and 
average avalanche size $\left\langle w \right\rangle$ for three different rebound parameters. Density $\rho=3/4$, 
the size of the lattice $N=7$.}
\label{tab:N7av}
\end{table}

\begin{table}[b]%
\begin{tabular}{c|c|c|c}
  & ${\mu_i}=const.$ &  $\mu_i=\delta/i$ & $\mu_i=\sigma/i^2$ \\
\hline
 $R={t_{L \rightarrow L}}/{t_{1 \rightarrow 1}}$ &    $\approx1.4844$    & $\approx1.6887$   &  $\approx 2.7001$     
\end{tabular}
\caption{ Coefficient $R={t_{L \rightarrow L}}/{t_{1 \rightarrow 1}}$  for three different rebound parameters (see main text for details).
Density for all cases $\rho=3/4$, the size of the lattice $N=7$.}
\label{tab:N7tavR}
\end{table}

For $N=7$ we investigate properties of the system with the density $\rho=3/4$.
Parameters are chosen as follows:
$\mu=1$,   $\nu=173024/10000$ gives the density $\rho=0.7500001621$,
$\theta=1/3$ gives $\rho=3/4$ exactly, 
and $\mu=1$, $\nu:=1000000/1578886$ gives $\rho=0.7500002817$.
Distributions of clusters are presented in Figure \ref{fig:Fig3a} and average cluster and avalanche sizes in Table \ref{tab:N7av}.
Again differences in distributions $n_i$ are not big, but average avalanche size differs significantly between considered cases.  

The novel property visible in the figure is that the highest probability is for the cluster of maximal size $i=N$. 
Thus, the system prefers merging clusters for high density.

A comparison of the ratios of return times $R={t_{L \rightarrow L}}/{t_{1 \rightarrow 1}}$ is presented in the right panel of Figure \ref{fig:FigRet5}, while formulas are presented in the Appendix. In each case the ratio is a rational function of $t$, which is equal to $1$ for $t=0$ and asymptotically $\sim t^6$ for $t\longrightarrow \infty$, which supports a Conjecture formulated in Section \ref{sec:conclusions}.
Table \ref{tab:N7tavR} shows that for the cases discussed above, with average density $\rho=3/4$, the highest value of
the ratio $R$ is  for $\mu_i=\delta/i$ and the smallest for $\mu_i=const.$
(which does not differ much from the value for $\mu_i=\delta/i$).   This is an opposite order comparing to the case with $\rho=1/5$ for $N=5$ considered above.
Thus, for higher densities the automaton prefers more periodic-like behaviour when it is relatively easier to trigger big avalanches. 

\begin{figure}[t]  
	\centering
	\includegraphics[width=8cm]{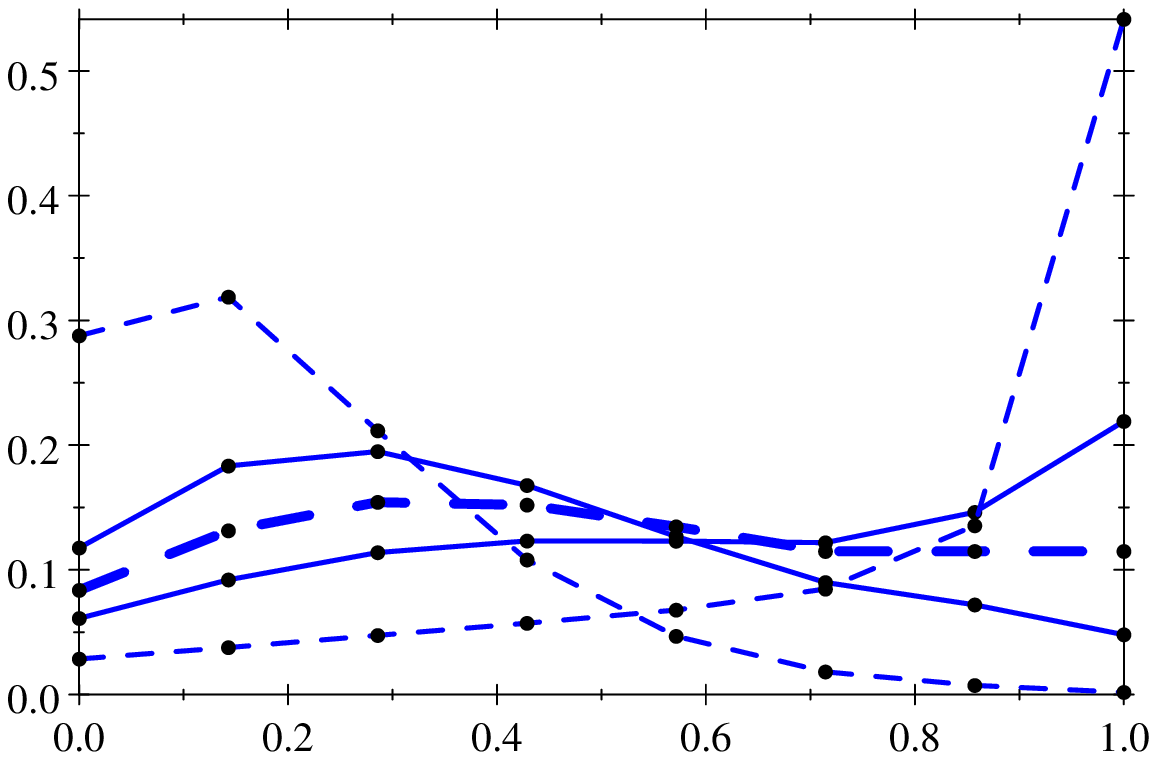} 
	\caption{Probability distributions of actual density of the system for the case $\mu_i/\nu=\theta/i$ for various average densities of $0.2,0.4,0.5,0.6,0.8$ -- the respective parameters are $\theta=4,\frac{3}{2},1, \frac{2}{3},\frac{1}{4}$.  On the plot: smaller average density corresponds to the higher probability rate for density equal to $0$ (and to the lower probability rate for density equal to $1$).
	The size of the system is $N=7$.} \label{fig:Fig3d}
\end{figure}

The size $N=7$ is big enough to notice how the actual density of the system 
(possible values are $0,\frac{1}{7},\frac{2}{7},\frac{3}{7},\frac{4}{7},\frac{5}{7},\frac{6}{7},1$) 
is distributed for various average densities. Results are shown in Figure \ref{fig:Fig3d}.
For small densities, like $\rho=0.2$, the maximum is for small $i$, that means that big densities and big avalanches are rare.
Then, when the density increases, the bell-like shape distribution appears and its maximum is shifted to the bigger values. 
Next, for densities like  $0.6$ or bigger, the 
maximum probability is for biggest possible size $i=N$ and the most probable state is that with $\rho=1$. 
To achieve big average density, the system must spend a substantial time being fully occupied.
The evolution of such a system consists of two phases: filing up and waiting for avalanche of maximal size, as is 
described above while discussing the times of return for $N=3$. 

For $N=500$ and constant parameters $\mu=\nu=1$, numerical experiments show that the density fits a Gaussian distribution \cite{BiaCz-Mon}.

\subsection{$N=10$}

In the example with the biggest $N$ presented here we investigate in several cases influence of correlations and compare  exact results with 
proposed approximations for $\alpha^A_{i}$, $\gamma^A_{i}$ and $\gamma^{AR}_{i}$. On the other hand, size $N=10$ requires relatively 
complex calculations -- 
the transition matrix is of size $108 \times 108$ and has about $1000$ non-zero entries.
All possible states are presented in Table \ref{tab:statesN10}.

The size $N=10$ is the smallest with states which consist of the same clusters, but in essentially different order. 
(For smaller $N$ states with different order of clusters were equivalent with respect to reflections.)
Namely, the state $88$ 
\begin{displaymath}
	\hookrightarrow | \ \ \ | \bullet | \ \ \ | \bullet | \ \ \ | \bullet | \bullet | \ \ \ | \bullet | \bullet | \hookleftarrow
\end{displaymath}
and the state $89$ 
\begin{displaymath}
	\hookrightarrow 	| \ \ \ | \bullet | \ \ \ | \bullet | \bullet | \ \ \ | \bullet | \ \ \ | \bullet | \bullet | \hookleftarrow
\end{displaymath}
In this subsection we consider also the relative difference between probabilities of these two states, namely $\Delta=(p_{88}-2p_{89})/p_{88}$ for various rebound parameters as a measure of adequacy of  independence of clusters assumption. 
 The multiplier $2$ in the above formula is necessary because the multiplicity of state $89$ is equal to five, and the multiplicity of the state $88$ is equal to ten. This quantity reflects dependence of respective probabilities on specific order of clusters in the system.  We assume there is no such dependence in order to write down approximations $\alpha^A_{i}$, $\gamma^A_{i}$ and $\gamma^{AR}_{i}$.

Other quantities analysed in examples below are $\frac{\alpha^E_{1}-\alpha^A_{1}}{\alpha^E_{1}}$,  
 $\frac{\alpha^E_{4}-\alpha^A_{4}}{\alpha^E_{4}}$,
 $\frac{\gamma^E_{5} - \gamma^A_{5}}{\gamma^E_{5}}$,
 $\frac{\gamma^E_{5} - \gamma^{AR}_{5}}{\gamma^E_{5}}$,
 $\frac{\gamma^E_{9} - \gamma^A_{9}}{\gamma^E_{9}}$ and
 $\frac{\gamma^E_{9} - \gamma^{AR}_{9}}{\gamma^E_{9}}$.
These quantities measure the quality of approximation formulas for $i=1$ and $4$ for $\alpha$ coefficients, and for 
$i=5$ and $9$ for $\gamma$ coefficients - just to test approximations for $n_2$ -- the first approximate equation, $n_5$ 
 -- the middle one, and $n_9$ the last one (those for $n_1$ and $n_{10}$ are exact).
A formula for exact value of $\alpha^E_{4}$, obtained from detailed analysis of states of the automaton, 
is presented in Appendix.

\begin{figure}[t]  
	\centering
	\includegraphics[width=4cm]{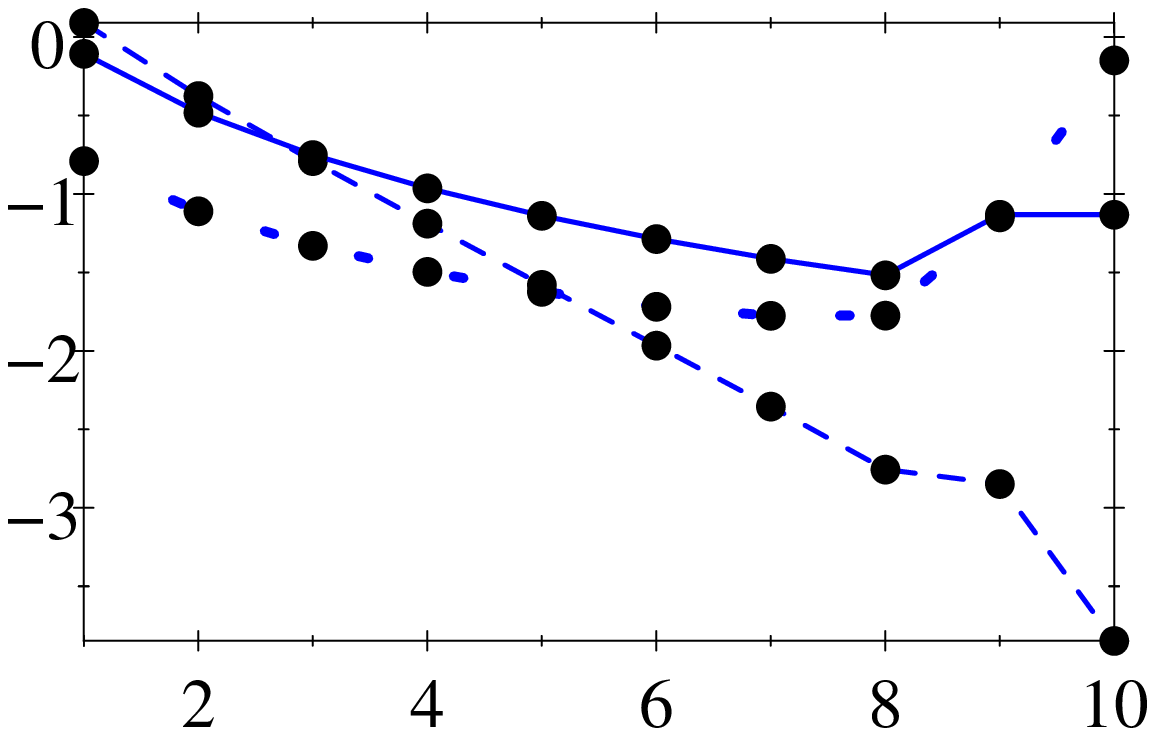} \quad 
	\includegraphics[width=4cm]{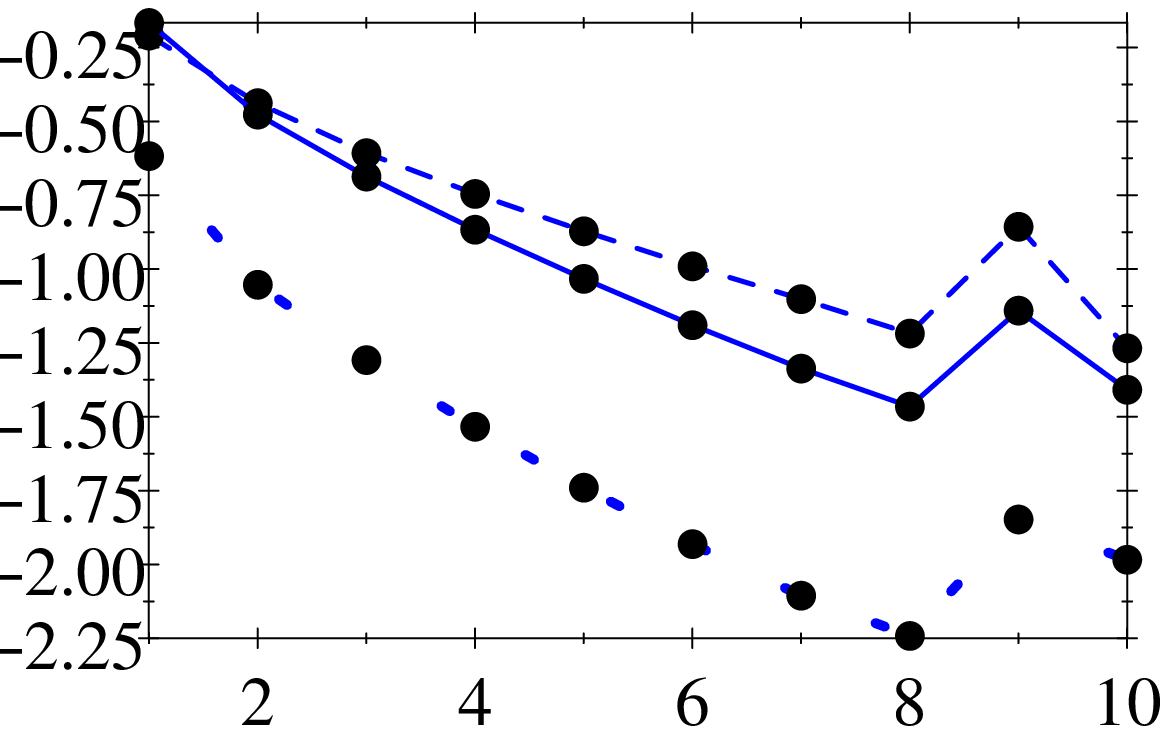} 
	\caption{Plot of the $Log_{10}$ of $n_i$s (left) and $n_i^0$ (right) for $N=10$ in three cases:  ${\mu_i}=const.$  
	(dashed line), $\mu_i=\delta/i$ (solid line) and  $\mu_i=\sigma/i^2$ (dotted line).
	} \label{fig:Fig10_3}
\end{figure}

\begin{table}[b]%
\begin{tabular}{c||c|c|c}
  & ${\mu_i}=1$ &  $\mu_i=1/i$ & $\mu_i=1/i^2$ \\
\hline
$\rho$ & 0.3076370614 & 0.5  & 0.8822697788\\
 $\left\langle i \right\rangle$ &    $1.5985438$    & $2.872872532$   &  $7.493995763$ \\
 $\left\langle w \right\rangle$ &    $2.250583644 $    & $2.872872532$   &  $3.725820785$   \\
 $\frac{p_{88}-2p_{89}}{p_{88}}$ &   $0.00865$  &  $0.01899$  &   $0.01868$ \\
 $\frac{\alpha^E_{1}-\alpha^A_{1}}{\alpha^E_{1}}$ &  $0.00909$  &  $0.066784$  &  $0.19468$ \\
 $\frac{\alpha^E_{4}-\alpha^A_{4}}{\alpha^E_{4}}$ &  $0.08795 $ & $0.07428$ & $0.09239 $ \\
 $\frac{\gamma^E_{5} - \gamma^A_{5}}{\gamma^E_{5}}$ &  $-0.01141$  &  $-0.28676$  &  $-0.89662$  \\
 $\frac{\gamma^E_{5} - \gamma^{AR}_{5}}{\gamma^E_{5}}$ &  $-0.00110$  &  $-0.06842$  &  $-0.13292$  \\
 $\frac{\gamma^E_{9} - \gamma^A_{9}}{\gamma^E_{9}}$ &  $0.35717$  &  $0.29878$  &  $0.17045$  \\
 $\frac{\gamma^E_{9} - \gamma^{AR}_{9}}{\gamma^E_{9}}$ &  $0.48970$  &  $0.62573$  &  $0.71428$ \\
 $t_{1\rightarrow 1}$  & 18.51 & 25.59 & 96.28\\
 $t_{L\rightarrow L}/t_{1\rightarrow 1}$& 379.61 & 5.2988 &  1.4606
\end{tabular}
\caption{Three cases: $\mu=1$, $\delta=1$ and $\sigma=1$ (and always $\nu=1$) for the size of the lattice $N=10$.}
\label{tab:N10-3}
\end{table}

\paragraph{Cases with constants equal to $1$.}

As a first set we consider three cases with the minimal possible rebounds factors, i.e., we put all constants equal to $1$.
Cases with $\mu_i=1$, $\mu_i=1/i$ and $\mu_i=1/i^2$ with $\nu=1$ are presented in Figure \ref{fig:Fig10_3} and Table \ref{tab:N10-3}.

Three different rebound parameter types result in various average density values, and hence different distributions. 
In all cases, the assumption of independence of clusters is well satisfied; the respective error $\Delta$ does not exceed $2\%$. An approximation for $\alpha^E_{4}$ is less than $10\%$ for all cases, but $\alpha^E_{1}$ strongly depends on the case (in fact it depends on density, as will be seen below). Approximation formulas for $\gamma_E$ perform in diversified way -- $\gamma_{AR}$ is better for mid $i$ terms,
while $\gamma_A$ is better for big $i$ terms. Nevertheless, both cases provide rather roughly appropriate values. 
These examples also suggest that for higher densities the system exhibits a periodic-like evolution.

\paragraph{Big densities}


\begin{figure}[t]  
	\centering
	\includegraphics[width=4cm]{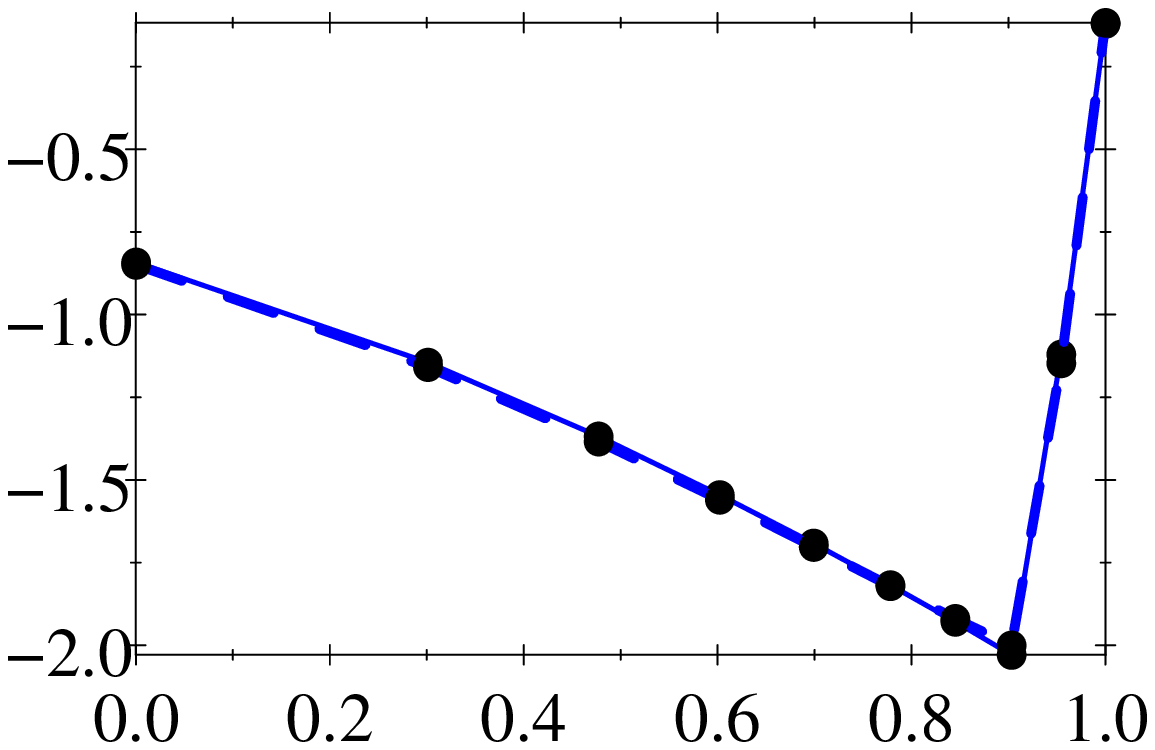} \quad 
	\includegraphics[width=4cm]{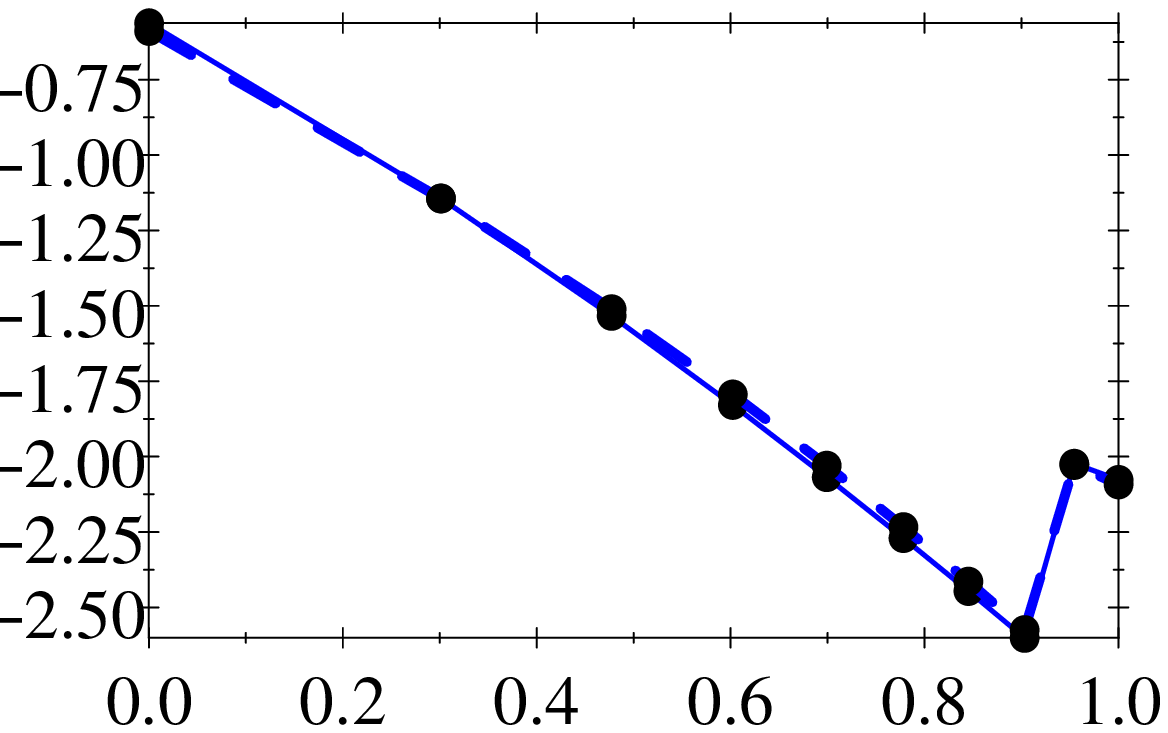} \\
	\includegraphics[width=4cm]{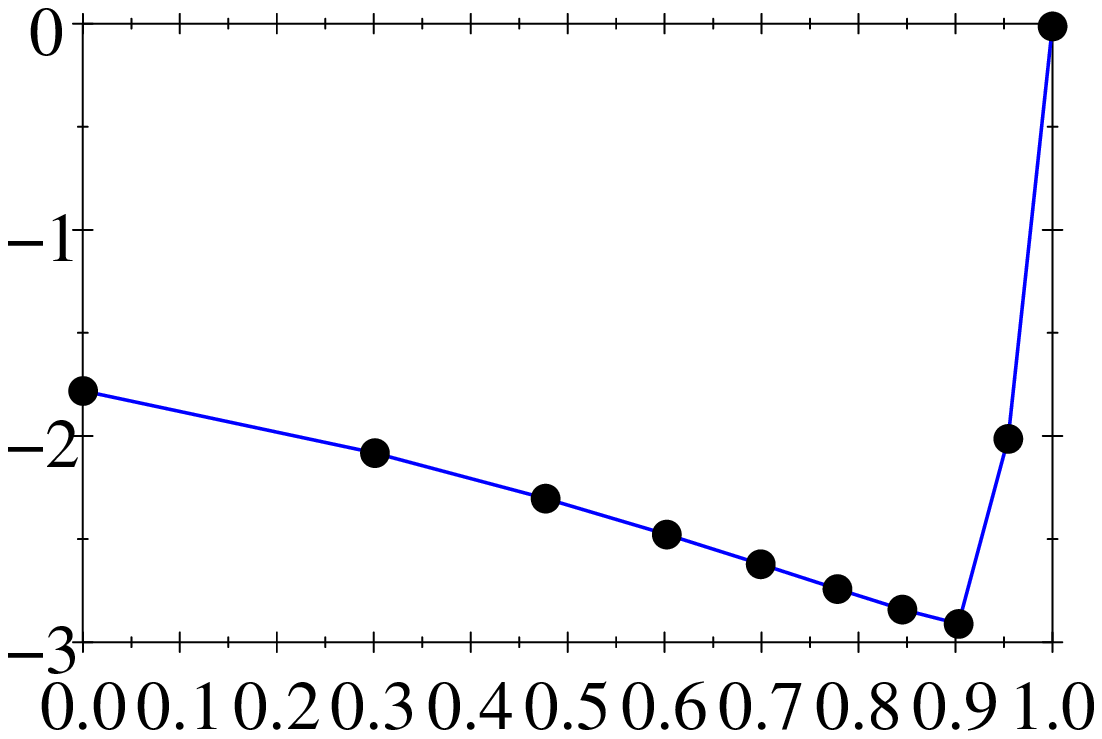} \quad 
	\includegraphics[width=4cm]{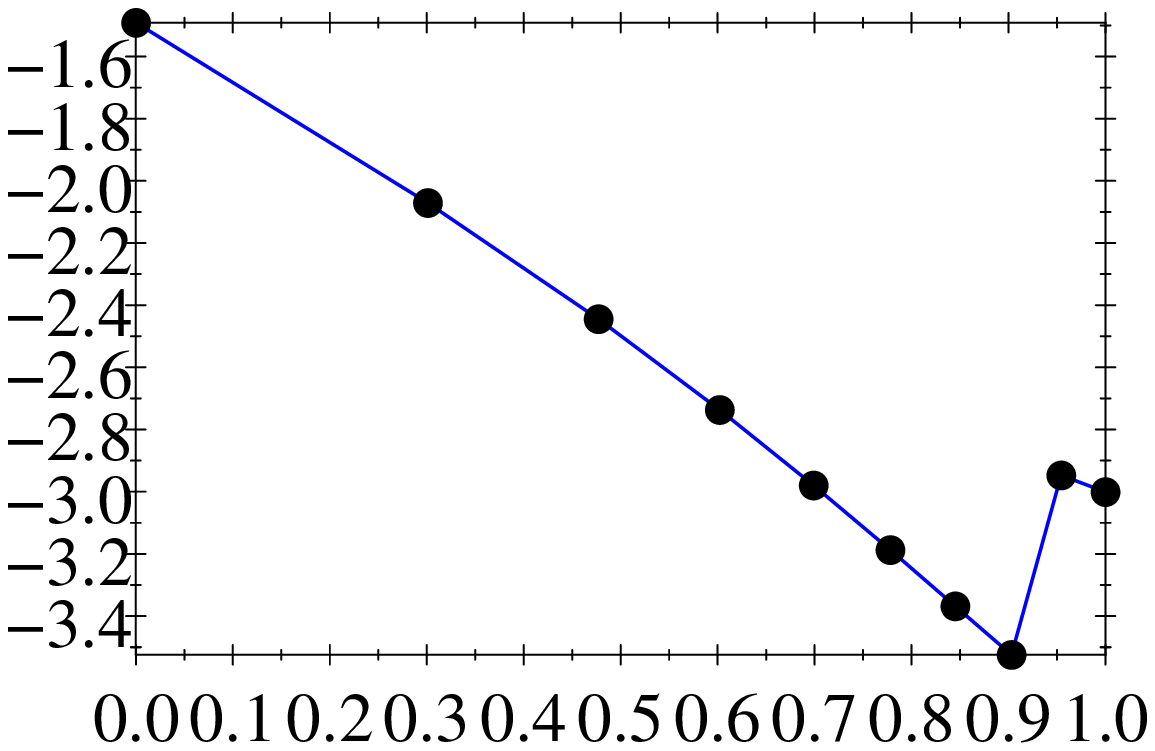} 
	\caption{Plot of the $Log_{10}$ of $n_i$s and $n_i^0$ vs. $Log_{10}(i)$ for $N=10$ in two cases:  ${\mu_i}=const.$  
	(solid line) and $\mu_i=\delta/i$ (dashed line) -- upper panels, and in case $\mu_i=\delta/i^2$ -- lower panels.
	} \label{fig:Fig10_big_rho_LL}
\end{figure}

\begin{table}[b]%
\begin{tabular}{c||c|c|c}
  & ${\mu_i}\sim1$ &  $\mu_i\sim1/i$ & $\mu_i\sim1/i^2$ \\
\hline
$\rho$ & 0.9145269069 & 0.9145269069 &  0.9897960692\\
 $\left\langle i \right\rangle$ &  7.764600567  & 7.805017612& 9.700144892\\
 $\left\langle w \right\rangle$ &  9.346154002  & 7.805017612& 8.33021261\\
 $\frac{p_{88}-2p_{89}}{p_{88}}$ & 0.00161 &0.00464& 0.00191\\
 $\frac{\alpha^E_{1}-\alpha^A_{1}}{\alpha^E_{1}}$ & 0.30898 &0.28421& 0.31377\\
 $\frac{\alpha^E_{4}-\alpha^A_{4}}{\alpha^E_{4}}$ &  -0.01511 & 0.01520 & 0.00330 \\
 $\frac{\gamma^E_{5} - \gamma^A_{5}}{\gamma^E_{5}}$ &  -0.96763  &  -0.93825  &  -1.05131  \\
 $\frac{\gamma^E_{5} - \gamma^{AR}_{5}}{\gamma^E_{5}}$ &  -0.15298  &  -0.14354  &  -0.14789 \\
 $\frac{\gamma^E_{9} - \gamma^A_{9}}{\gamma^E_{9}}$ & 0.42749 & 0.39218& 0.38965 \\
 $\frac{\gamma^E_{9} - \gamma^{AR}_{9}}{\gamma^E_{9}}$ & 0.80211 &  0.78888 & 0.80019 \\
 $t_{1\rightarrow 1}$  & 119.18 & 123.60\footnote{The system stays in fully occupied state 
 $1/\mu_{10}\approx107,5$, which is longer than $100$ as in $\mu=const.$ case (previous column). The respective 
 average times for filling up the lattice are $\approx 19$ and $\approx16$. } & 1002.44 \\
 $t_{L\rightarrow L}/t_{1\rightarrow 1}$ &  1.1066 & 1.1339 & 1.0277
\end{tabular}
\caption{Three cases with "big" $\rho$ for the size of the lattice $N=10$ (see main text for details).}
\label{tab:N10-bigrho}
\end{table}


In order to investigate evolution of the system with high average density 
(and strong deviations in actual density) we consider
case $\mu=const$ with $\mu_1=1/100$ and $\nu=1$, which gives the density $\rho\approx 0.91$, and  
case $\delta/i$ with $\mu_1=4673077001/5*10^{10}\approx0.093$ and $\nu=1$ to obtain the same density (with $10$ digits accuracy)
for comparison. 
Also we consider case of $\sigma/i^2$ with $\mu_1=1/10$ and $\nu=1$ which gives the density $\rho\approx 0.99$.
The results are presented  in Figure \ref{fig:Fig10_big_rho_LL} and Table \ref{tab:N10-bigrho}.

Plots of respective distributions for $\mu_i\sim1/i$ and ${\mu_i}\sim1$ are overlapping each other. For relatively small size $N=10$,
fixing the average density of the system strongly determines distributions, making the dependence on rebound parameters not essential.
Their influence becomes more visible for larger sizes $N$ of the lattice. 
In case of high density, the system just spend much time being fully occupied.  

For high densities, the assumption of independence of clusters is well satisfied; the respective error $\Delta$ does not exceed $0.5\%$. 
 An approximation for $\alpha^E_{4}$ is fairly good ($\approx 1.5\%$ or less), but $\alpha^E_{1}$ has only accuracy $\approx 30\%$. Approximation formula for $\gamma_{AR}$ is much better for mid $i$ terms (though giving only $\approx 15\%$ accuracy),
while $\gamma_A$ is better for big $i$ terms ($\approx40\%$). 
Thus, for high density cases the proposed set of equations for $n_i$s does not reproduce actual distribution.
Note, however, that there are other exact equations valid for any density.
 
The parameter $t_{L\rightarrow L}/t_{1\rightarrow 1}$ for  $\mu_i\sim1/i$ case is bigger than for ${\mu_i}\sim1$  case (both cases have the same "big" density), which agrees with the results for $N=7$ with $\rho=3/4$ presented in Table \ref{tab:N7tavR}.

\begin{table}[b]%
\begin{tabular}{c||c|c|c}
  & ${\mu_i}\sim1$ &  $\mu_i\sim1/i$ & $\mu_i\sim1/i^2$ \\
\hline
$\rho$ & 0.0779280356 & 0.01031150521&  0.01031150521\\
 $\left\langle i \right\rangle$ & 1.09149321 & 1.02083788 & 1.041894016\\
 $\left\langle w \right\rangle$ & 1.183235221  &1.02083788& 1.020408163\\
 $\frac{p_{88}-2p_{89}}{p_{88}}$ & $6.9231*10^{-5}$ & $6.4609*10^{-4}$ & $3.5663*10^{-3}$\\
 $\frac{\alpha^E_{1}-\alpha^A_{1}}{\alpha^E_{1}}$ & $1.7483*10^{-5}$ & $2.7693 *10^{-7}$& $1.0585*10^{-6}$\\
 $\frac{\alpha^E_{4}-\alpha^A_{4}}{\alpha^E_{4}}$ &  $0.14229$ & $0.16313$ & $0.16274$ \\
 $\frac{\gamma^E_{5} - \gamma^A_{5}}{\gamma^E_{5}}$ &  $0.0009069$  &  $0.0020519158$  &  $0.0040100$  \\
 $\frac{\gamma^E_{5} - \gamma^{AR}_{5}}{\gamma^E_{5}}$ &  0.0009101  & 0.0020519262  &   0.0040124 \\
 $\frac{\gamma^E_{9} - \gamma^A_{9}}{\gamma^E_{9}}$ & 0.09615 &0.01473& 0.00606 \\
 $\frac{\gamma^E_{9} - \gamma^{AR}_{9}}{\gamma^E_{9}}$ & 0.12138& 0.02356 &  0.03305 \\
 $t_{1\rightarrow 1}$  & 22.32 & 106.35 & 110.57 \\
 $t_{L\rightarrow L}/t_{1\rightarrow 1}$ & 2220903488.0  &  $1.666292752\cdot10^{14}$ & 9971770329.0 \\
                                          &  $\sim 2\cdot 10^{10} $  &  $\sim 2\cdot 10^{14} $ &  $\sim 1\cdot 10^{11} $
\end{tabular}
\caption{Three cases with "small" $\rho$ for the size of the lattice $N=10$ (see main text for details).}
\label{tab:N10-smallrho}
\end{table}

\paragraph{Small densities}

\begin{figure}[tb]  
	\centering
	\includegraphics[width=4cm]{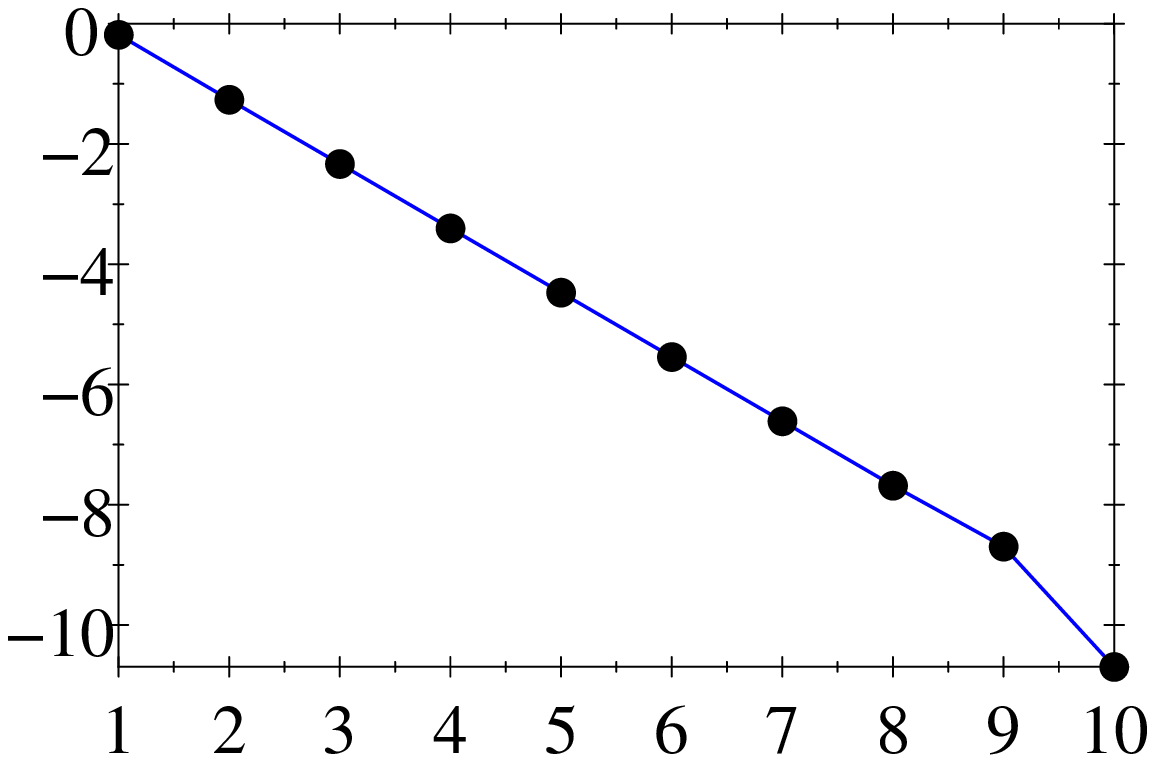} \quad 
	\includegraphics[width=4cm]{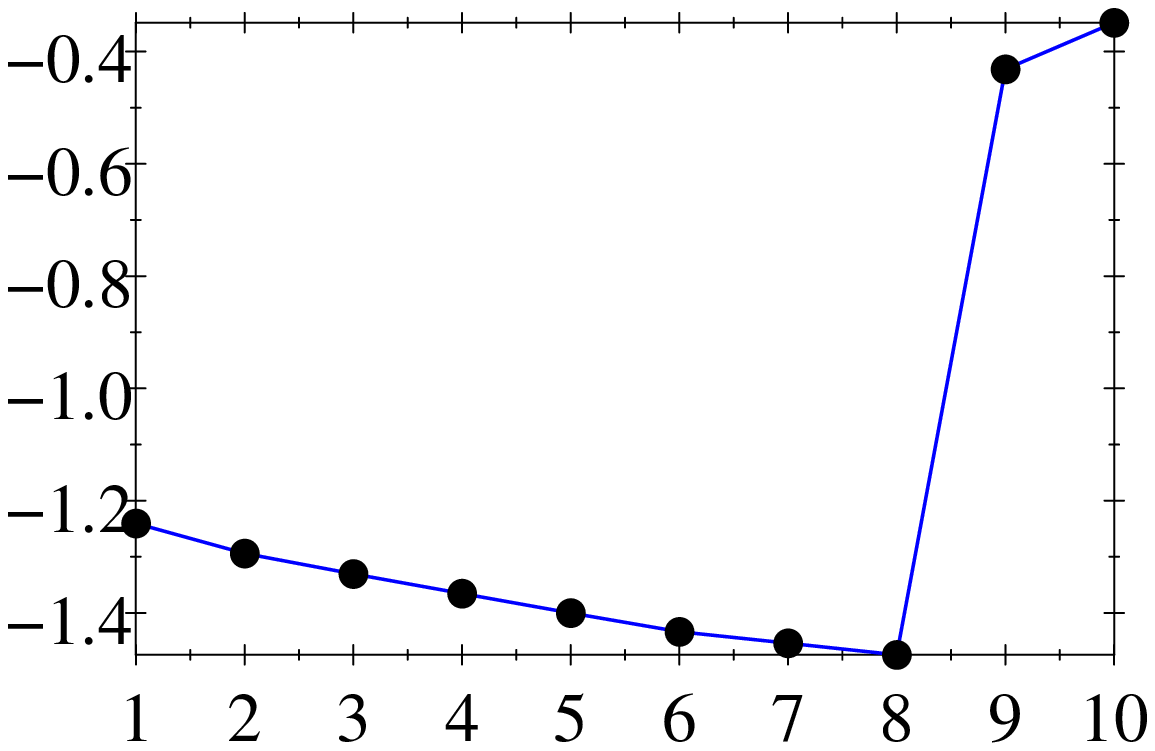} \\
	\includegraphics[width=4cm]{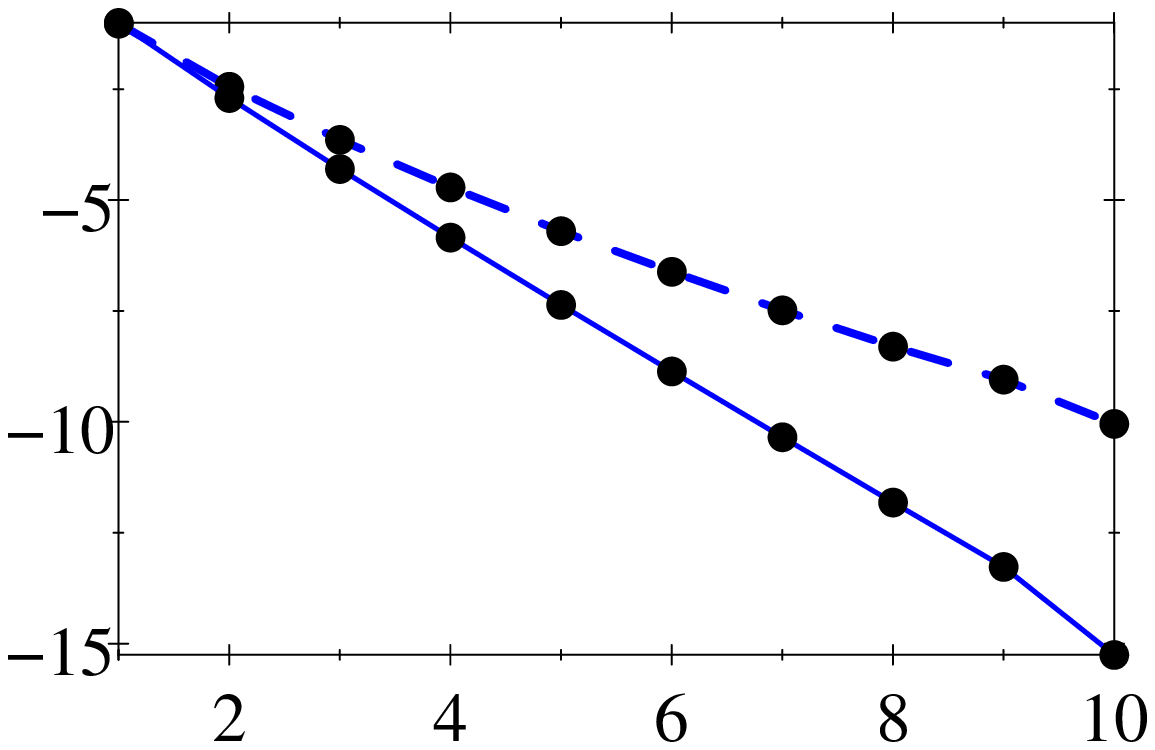} \quad 
	\includegraphics[width=4cm]{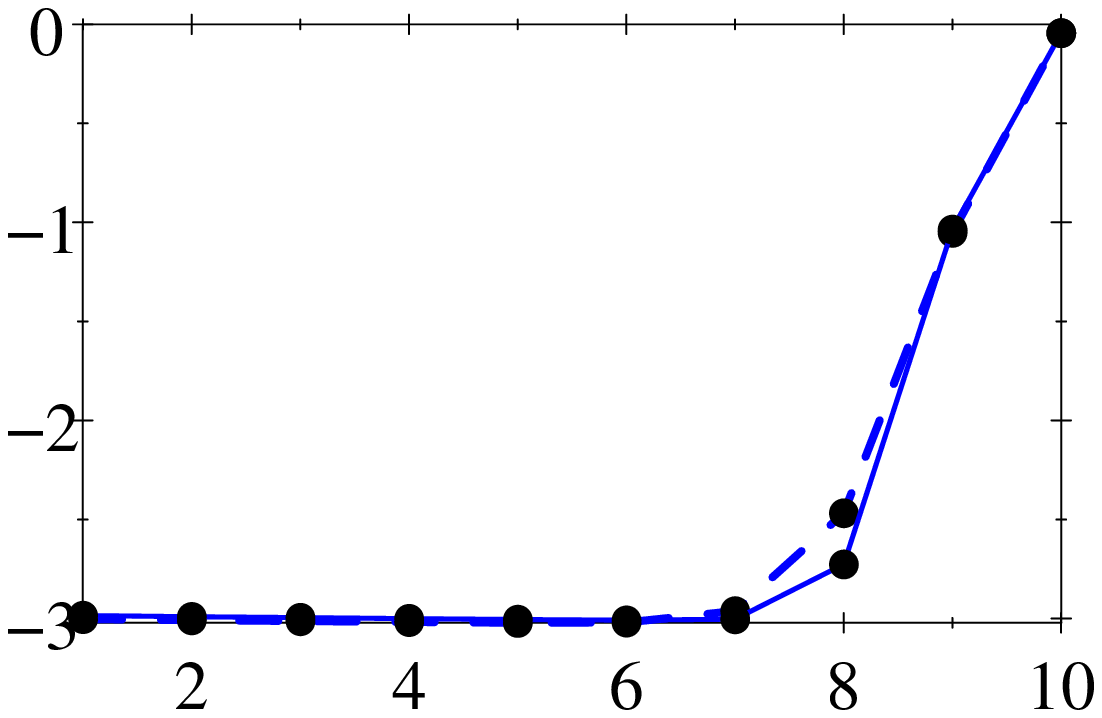} 
	\caption{Plot of the $Log_{10}$ of $n_i$s and $n_i^0$ for $N=10$ in  case  ${\mu_i}=const.$ -- upper line, 
	and in two cases:  ${\mu_i}=\delta/i$  (solid line) and $\mu_i=\delta/i^2$ (dashed line) -- lower line.
	} \label{fig:Fig10_small_rho}
\end{figure}

To present system behaviour in small average density we choose
$\mu_1=1$ and $\nu=1/10$ for case $\mu=const$ - it gives density $\rho\approx0.08$. 
Then for the remaining two cases we have the same density $\rho\approx0.01$ (with $10$ digits accuracy), 
with the following parameters:
 $\mu_1=1$ and $\nu=50000000/4798952601\approx 0.01$ - for case $\delta/i$  and  
$\mu_1=1$ and $\nu=1/100$ for case $\sigma/i^2$. The results are presented  in Figure \ref{fig:Fig10_small_rho} and Table \ref{tab:N10-smallrho}.

For small densities assumption of independence of clusters is well satisfied.  In general, all proposed approximations are 
fairly good.
 An approximation for $\alpha^E_{4}$ is the worst; its accuracy is only $\approx 15\%$.
 As previously, approximation formula for $\gamma_A$ is better than $\gamma_{AR}$ for big $i$ terms, but
it appears that for mid $i$ terms both formulas give almost the same values (because $n_i$s decrease rapidly).
Thus, for small densities the set of equations for $n_i$s can be used to reproduce the actual distribution.

It is very improbable to find the lattice fully occupied for small average densities, which is reflected in high values of the parameter $R=t_{L\rightarrow L}/t_{1\rightarrow 1}$.  
The parameter $R$ for  $\mu_i\sim1/i$ case is bigger than its for  ${\mu_i}\sim1/i^2$  case (both cases have the same "small" density), which agrees with the results for $N=5$ with $\rho=1/4$ presented in Table \ref{tab:N5tavR}.

\section{Conclusions} \label{sec:conclusions}

In this article we investigated in detail a finite version of one-dimensional non-equilibrium dynamical system -- Random Domino Automaton. 
It is a simple, slowly driven system with avalanches. The advantage of RDA (comparing to Drossel-Schwabl model) is
the dependence of rebound parameters on the size of a cluster. 
This crucial extension allows for producing a wider class of distributions 
by the automaton, as well as leads to several exact formulas.
Exponential type and inverse-power type distributions of clusters were studied in \cite{BiaCzAA}; the present work
examines also V-shape distributions and quasi-periodic like behavior. 
 
Detailed analysis  of finite RDA, including finite size effects, 
extends and explains the previously obtained results for RDA. Moreover, we also analyzed approximations made when deriving  
equations for the stationary state of the automaton. This allows for the following conclusions.

The balance of $\rho$ equation \eqref{eq:rho_balance} and the balance of $N$ equations \eqref{eq:baln} are exact 
-- their forms incorporate all correlations present in the system. The first one has a form independent of the 
size of the lattice $N$, thus it is exactly the same as for RDA.  The second one contains correction for finite size effect,
namely a term $(2n_R-n_{N-1})$, which replaces the term $2n$ for RDA. When $n_{N-1}$ and $n_N$ are negligible, these two terms coincide.  
For finite RDA, balance of $n_i$s equations \eqref{eq:baln1}-\eqref{eq:balnN} contains  two extra equatins, for $i=N-1$ and $i=N$, comparing to the those for RDA.  The first (for $n_1$) and the last (for $n_N$) are exact. Note that all those equations are written for 
rebound parameter $\mu=\mu(i)$ being a function of cluster size and $\nu$ being a constant.

The most remarkable special case is when $\mu=\delta/i$, when any cluster has the same probability to be removed as an avalanche independently of its size $i$. It appears that the system depends on a single parameter $\theta=\delta/\nu$, or 
equivalently, due to neat exact formula (eq. \eqref{eq:densityB}) 
\begin{equation*}
\rho=\frac{1}{1+\theta},
\end{equation*}
the properties of the system may be characterized by the value of the average density.
Note that the above expression does not depends on the size $N$, and is the same as for RDA. 
This specialization leads to more neat formulas, like the 
equation  for $n_R$ (eq. \eqref{eq:nRB})
\begin{equation*}
n_R=N\frac{\theta}{(\theta+1)(\theta+2)}.
\end{equation*}
Note again that it has the same form as for RDA, except that $n$ is replaced by $n_R$ ($n=n_R+n_N$). 
Summarizing, the model allows  to derive a number of explicit dependencies, as shown in Sections \ref{sec:equations} and \ref{sec:cases}. 

The Random Domino Automaton defines a discrete time Markov process of order 1 and, in principle, may be solved exactly.
However, it turns out that computations are fairly complex  and exact formulas are long, as visible from examples presented in the Appendix. 
Also, the exact numerical values are in the form of big numbers --- in every considered example ($N=3,4,5,6,7,10$) significantly big prime numbers were encountered. 
For example, for the simplest possible rebound parameters ($\mu=1$ and $\nu=1$) the exact value of denominators of probabilities of states for $N=10$ (see Table \ref{tab:statesN10}) is a 65-digit integer. Its prime factorization (presented in the Appendix)
contains a 56-digit integer, which cannot be simplified with numerators.
Thus, the usefulness of Markov chains for finding both formulas and values $n_i$s is limited in practice.

Nevertheless, Markov chains framework leads to interesting results concerning analysis of times of recurrence for specific states. 
A return time  to the state with density $\rho=1$ (equation \eqref{eq:rettimeL})
\begin{equation*}
t_{L \rightarrow L}= 1/\mu_N + \hat\tau_1,
\end{equation*}
consists of two parts: waiting of fully occupied lattice for triggering a maximal avalanche and "loading" time, when the lattice is filled up, respectively.
If the average density of the system is small, the second time is very long. 
The formula is more interesting  for systems with relatively big average density, when the "loading" time is comparable to waiting time for triggering the biggest avalanche. Such a system exhibits a periodic like behavior. 
Dividing the waiting time $t_{L \rightarrow L}$ by the waiting time $t_{1 \rightarrow 1}$ (given by equation \eqref{eq:rettime11})
one has the following measure of quasi-periodicity
$$ R=  t_{L \rightarrow L} / t_{1 \rightarrow 1} =  t_{1 \rightarrow L \rightarrow 1} / t_{1 \rightarrow 1}. $$
If $R=1$ then the system is periodic.

Several considered examples lead to the following conjecture concerning the coefficient $R$.  \\
{\bf Conjecture.} The ratio of return times $t_{L\longrightarrow L}/t_{1\longrightarrow 1}$ as a function of $t$ being the ratio of constants from rebound parameters ($\mu/\nu$, $\delta/\nu$, $\sigma/\nu$) are rational functions of $t$, $f(t)=t_{L\longrightarrow L}/t_{1\longrightarrow 1}$ with the following properties for any size $N$ of the system
\begin{eqnarray}
f(t=0)&=& 1, \\
\lim_{t \rightarrow \infty} \frac{f(t)}{t^{N-1}} &=& const.
\label{eq:conj}
\end{eqnarray} 
The conjecture relates the size of the system $N$ with asymptotic behavior of ratio of waiting times. 

There are big fluctuations (variations of actual density) during the evolution of systems with relatively big average densities. 
If the system is likely to achieve a fully occupied state, the next state is an empty state, and the variations in density are maximal. Nevertheless, some parameters of stationary state (more precisely, statistically stationary state) satisfy exact equations,
as shown above. For big average densities, the system fluctuates within the whole possible range, and cannot be thought of as having approximately stationary values during the evolution.  
This aspect is easy to be overlooked (see \cite{PBfire}).

It is argued in the Appendix that no exact equations for $n_i$s exist for the size $N \geq 5$.
Thus, to have compact equations for $n_i$s, some approximation formulas are proposed. 
The first general conclusion from the examples is that the approximations are acceptable for small densities, but 
for big densities the errors are substantial. 
The main reason is that for big densities correlations become more important and fluctuations makes actual values of the parameters substantially different from their stationary values, which are present in the formulas.
These properties are particularly severe for small sizes of the system, where every avalanche changes the actual density considerably. 

Table \ref{tab:comp_alpha} presents a dependence of a relative error of $\alpha^E_{1}$ with respect to $\alpha^A_{1}$ on size $N$ of the system. For bigger $N$ the accuracy of approximation is growing, which corresponds well with the remark in the last paragraph.
\begin{table}[t]
\begin{tabular}{c|c|c|c|c}
$N$  & $\alpha^E_{1}=\frac{2n_2}{n_1}$ & $\alpha^A_{1} = \left(1-\frac{n_1^0}{n_R} \right)$ & $\frac{\alpha^E_{1} - \alpha^A_{1}}{\alpha^E_{1}}$ & $\rho$ \\
\hline
$3$  &   $ 1.33(3) $ &  $0.6$  & $0.55$ & $\approx 0.3462$\\
$4$  &   $ 0.66(6) $ &  $\approx 0.6316 $  & $\approx 0.053 $ & $0.32(32)$\\
$5$  &   $ \approx 0.6829 $    &  $\approx 0.6565$  & $\approx0.039$ & $\approx 0.3139$\\
$6$  &   $ \approx 0.685296$    &  $\approx 0.669232$  & $\approx0.023$ & $\approx 0.3102$\\
$7$  &   $ \approx 0.685523$    &  $\approx 0.675066$  & $\approx0.015$ & $\approx 0.3086$\\
$10$  &   $\approx 0.685436  $    &  $\approx 0.679205$\footnote{$(1-{n_1^0}/{n})\approx0.679229$}  & $\approx0.0091$ & $\approx0.3076$ \\
$4000$\footnote{Results from \cite{BiaCzAA}} &  $\approx 0.677$  & $\approx 0.677$ &  --- & $\approx 0.3076$
\end{tabular}
\caption{Comparision of $\alpha^E_{1}$ with $\alpha^A_{1}$ for $N=3,4,5,6,7$ and $10$ for parameters $\mu=1$ and $\nu=1$.
The last line  presents value of $\alpha^A_{1}$ for $N=4000$ obtained from simulations and equations (respectively) in \cite{BiaCzAA}.}
\label{tab:comp_alpha}
\end{table}

It can be noticed from the distributions of $n_i$s  of examples presented above that all $n_i$s except of the last two 
(namely $n_{N-1}$ and $n_N$) are placed on one "regular" curve, while the last two deviate from it. It may be regarded as 
a (correction of) finite size effect. Also in the respective set of equations \eqref{eq:baln1}-\eqref{eq:balnN}, the last two (for $i-N-1$ and $i=N$) have a form different from the previous ones. 
Thus, neglecting the size restriction, which in fact ignores the last two equations, is justified when the deviations of the last two 
$n_i$s from the "regular" curve are not big. That happens for small densities. 

It appears also that for index $i$ in his middle range of values an approximation formula $\gamma_{AR}$ works better than $\gamma_{A}$, in spite of the fact that it looks to be more rough approximation. For distribution of $n_i$s vanishing rapidly (i.e., for small densities) both give comparable results. 

All this justifies the form of equations for $n_i$s presented in \cite{BiaCzAA} as valid for small densities. A detailed examination 
of the RDA for big densities requires further investigations.

This article explores properties of FRDA in order prepare to modeling of real data.    
In this context, among others, formulas for waiting times can be used. 
We ephasize also a formula \eqref{eq:timeavalanche}
\begin{equation*}
t_{av}= \frac{\left\langle w \right\rangle + 1}{1-P_r},
\end{equation*}
which relates  the measure of scattering (dissipation) of balls $P_r$ with the average size of avalanche  $\left\langle w \right\rangle$ and  average time between any two consecutive avalanches $t_{av}$, which are a priori measurable quantities.

The Random Domino Automaton proved to be a stochastic dynamical system with interesting mathematical structure.
It may be viewed as extension of Drossel-Schwabl model, and we showed that this is a substantial generalization with a wide
range of novel properties. 
W expect it can also be applied to natural phenomena, including earthquakes and forest-fires.
This is our aim for the future work. 

\section*{Appendix. Exact equations for $N=3,4$ and their non-existence for $n\geq5$}

For arbitrary size $N$, there are four exact equations: balance of $\rho$ - equation \eqref{eq:rho_balance},
balance of $n$  - equation \eqref{eq:baln}, for $n_1$ - equation \eqref{eq:baln1} and for $n_N$ - equation \eqref{eq:balnN}.

\noindent{\bf Size $N=3$.} 
Equation for $n_2$ is of the form \eqref{eq:balnN-1}, namely
$$ n_2 =\frac{1}{2\frac{\mu_1}{\nu}+1} (2 n_1 \alpha^E_{1} + n_1^0 \gamma^E_2).$$
In this case the only companion to single one-cluster is an empty two-cluster (see state $2$ in Tab.\ref{tab:N3}), hence
$$\alpha^E_{1}=1 \quad \text{and} \quad  \gamma^E_{2}=0.$$
Thus, we arrived at the exact form of the equation for $n_2$.

\noindent{\bf Size $N=4$.}
\begin{table}[t]%
\begin{tabular}{c|c|c|c}
 state number&  example\footnote{Other states differ by shifts.} & multiplicity & contrib. to   \\
\hline\hline
1 & $ \hookrightarrow | \ \ \ | \ \ \ | \ \ \ | \ \ \ | \hookleftarrow $	& 1 & $ n_4^0 $ \\
\hline
2& $ \hookrightarrow | \ \ \ | \ \ \ | \ \ \ | \bullet | \hookleftarrow $	& 4 & $n_1, n_3^0$ \\
\hline
3 & $ \hookrightarrow | \ \ \ | \ \ \ | \bullet | \bullet | \hookleftarrow $	& 4 & $n_3, n_2^0$ \\
\hline
4& $ \hookrightarrow | \ \ \ | \bullet | \ \ \ | \bullet | \hookleftarrow $	& 2 & $ n_1, n_1^0 $ \\
\hline
5 & $ \hookrightarrow | \ \ \ | \bullet | \bullet | \bullet | \hookleftarrow $	& 4 & $ n_3, n_1^0 $ \\
\hline
6 & $ \hookrightarrow | \bullet | \bullet | \bullet | \bullet | \hookleftarrow $	& 1 & $ n_4 $ 
\end{tabular}
\caption{States  for the size of the lattice $N=4$. }
\label{tab:N4}
\end{table}
All states of the automaton and their labels are presented in  Tab.\ref{tab:N4}.
Equation for $n_2$ is of the form \eqref{eq:baln2}
$$ n_2 = \frac{2}{2\frac{\mu_2}{\nu}+2}n_1\alpha^E_{1},$$
where to $\alpha^E$ contributes only state $2$, not state $4$. Hence 
$$\alpha^E_{1}=\frac{p_2}{p_2+2p_4}=\left(1-\frac{2p_4}{n_1}\right)= \left(1-\frac{n_1^0-n_3}{n_1}\right),$$
where $p_i$ is probability of state $i$.
Thus $\alpha^E_{1}$ is expressed as function of $n_i$s and $n_1^0$.
The equation  for $n_3$ is of the form \eqref{eq:balnN-1} 
$$ n_3=\frac{1}{3\frac{\mu_3}{\nu}+1}\left( 2n_2 \alpha^E_{2} + n_1^0 \gamma^E_{3} \right).$$
In this case 
$$\alpha^E_{2}=1,$$
because only state $3$ contributes. The state $4$ (and not state $5$) contributes 
to $\gamma^E_3$, therefore 
$$\gamma^E_3=\frac{2p_4}{2p_4+p_5}=\left(1-\frac{p_5}{n_1^0}\right) =\left(1-\frac{n_3}{n_1^0}\right).$$
This completes the task of writing exact equations for $N=4$.

\noindent{\bf Size $N=5$.}
States and their labels are presented in  Tab.\ref{tab:N5}.
In this case, the coefficients are as follows
$$\alpha^E_{2}=\frac{p_2+p_4}{n_1}, \quad \quad \gamma^E_3=\frac{p_4-p_6-n_4}{n_1^0}, $$

$$\alpha^E_{3}=\frac{p_3}{n_2}, \quad \quad \gamma^E_4=\frac{p_6-p_4-n_4}{n_1^0}. $$
Summing up the probabilities contributing to $n_1^0, n_1$ and $n_2$ one obtains
\begin{eqnarray}
n_1^0 &=& p_4+ 2p_6+ n_4, \label{eq:n10fromstates}\\
n_1 &=& p_2+ 2p_4+p_6,  \\
n_2 &=& p_3+ p_6. \label{eq:n2fromstates}
\end{eqnarray} 
The set cannot be solved for $p_2, p_3, p_4 ,p_6$. Since there are no more equations for those coefficients,
respective $\alpha$s and $\gamma$s cannot be expressed as  functions of $n_i$s only in an exact manner.

\noindent{\bf Sizes bigger than $5$.}
An argument for non-existence of exact set of equations  \eqref{eq:baln1}-\eqref{eq:balnN}, i.e., non-existence of
exact formulas for $\alpha^E_{i}$s and $\gamma^E_{i}$s as functions of $n_i$s and $n_1^0$ is based on the same impossibility 
of solving equations as presented above.

An increase of size of a grid $N$ by $1$ results in an increase of  the set of $n_1, n_2, \ldots$ by one and much bigger increase of 
the number of states.  
An analog of the set of equations \eqref{eq:n10fromstates}-\eqref{eq:n2fromstates} will contain much more probabilities of states $p_1, p_2, \ldots$, on the right hand side -- there will be more states containing $1$-clusters, $2$-clusters and so on, and contributing to $n_1,n_2,\ldots$ respectively. 
Thus, it is impossible to express those probabilities of states as functions of $n_1^0, n_1,n_2,\ldots$ only.
As a consequence, there are no general exact formulas for $\alpha^E_{i}$s and $\gamma^E_{i}$.

\begin{widetext}

\section*{Appendix. Formulas} 

The return times for $N=5$ for general values of the parameters $\mu_1,\mu_2,\mu_3,\mu_4,\mu_5$ and $\nu$ are 

\bigskip

$t_{1\rightarrow1} = \frac{1}{\nu} + 
 \frac{1}{\mu_5 (4\mu_4 + \nu)(3\mu_3 + 2\nu)(2mu_1^3 \mu_2 + 2\mu_1^3\nu + 4\mu_1^2\mu_2^2 + 17\mu_1^2\mu_2\nu + 13\mu_1^2\nu^2 + 14\mu_1\mu_2^2\nu + 43\mu_1\mu_2\nu^2 + 33\mu_1\nu^3 + 16\mu_2^2\nu^2 + 42\mu_2\nu^3 + 36\nu^4)}
\times   
(48\mu_1\nu^5 + 60\mu_2\nu^5 + 54\mu_3\nu^5 + 750\mu_5\nu^5 + 72\nu^6 + 8\mu_1^2\nu^4 + 8\mu_2^2\nu^4 + 20\mu_1\mu_2\nu^4 + 12\mu_1\mu_3\nu^4 + 24\mu_2\mu_3\nu^4 + 560\mu_1\mu_5\nu^4 + 760\mu_2\mu_5\nu^4
+ 720\mu_3\mu_5\nu^4 + 1560\mu_4\mu_5\nu^4 + 100\mu_1^2\mu_5\nu^3 + 160\mu_2^2\mu_5\nu^3 + 40\mu_1\mu_2^2\mu_5\nu^2 + 20\mu_1^2\mu_2\mu_5\nu^2 + 60\mu_1^2\mu_3\mu_5\nu^2 + 240\mu_1^2\mu_4\mu_5\nu^2 
+ 150\mu_2^2\mu_3\mu_5\nu^2 + 480\mu_2^2\mu_4\mu_5\nu^2 + 340\mu_1\mu_2\mu_5\nu^3 + 390\mu_1\mu_3\mu_5\nu^3 +
1280\mu_1\mu_4\mu_5\nu^3 + 645\mu_2\mu_3\mu_5\nu^3 + 1840\mu_2\mu_4\mu_5\nu^3 + 1800\mu_3\mu_4\mu_5\nu^3 + 240\mu_1\mu_2^2\mu_3\mu_4\mu_5 + 120\mu1^2\mu_2\mu_3\mu_4\mu_5 + 285\mu_1\mu_2\mu_3\mu_5\nu^2 + 60\mu_1\mu_2^2\mu_3\mu_5\nu + 30\mu_1^2\mu_2\mu_3\mu_5\nu + 960\mu_1\mu_2\mu_4\mu_5\nu^2 + 160\mu_1\mu_2^2\mu_4\mu_5\nu + 80\mu_1^2\mu_2\mu_4\mu_5\nu + 1320\mu_1\mu_3\mu_4\mu_5\nu^2 + 240\mu_1^2\mu_3\mu_4\mu_5\nu + 2100\mu_2\mu_3\mu_4\mu_5\nu^2 + 600\mu_2^2\mu_3\mu_4\mu_5\nu + 1140\mu_1\mu_2\mu_3\mu_4\mu_5\nu),$ 
 
\bigskip
 
$t_{L\rightarrow L} =\frac{1}{\mu_5} + \frac{1}{2\nu^5(4\mu_1^2 + 10\mu_1\mu_2 + 24\mu_1\nu + 6\mu_3\mu_1 + 4\mu_2^2 + 30\mu_2\nu + 12\mu_3\mu_2 + 36\nu^2 + 27\mu_3\nu)}
\times
(626\mu_1\nu^5 + 844\mu_2\nu^5 + 828\mu_3\nu^5 + 1848\mu_4\nu^5 + 822\nu^6 + 126\mu_1^2\nu^4 + 4\mu_1^3\nu^3 + 192\mu_2^2\nu^4 + 8\mu_1^2\mu_2^2\nu^2 + 426\mu_1\mu_2\nu^4 + 489\mu_1\mu_3\nu^4 + 1544\mu_1\mu_4\nu^4 + 771\mu_2\mu_3\nu^4 + 2176\mu_2\mu_4\nu^4 + 2232\mu_3\mu_4\nu^4 + 68\mu_1\mu_2^2\nu^3 + 54\mu_1^2\mu_2\nu^3 + 4\mu_1^3\mu_2\nu^2 + 99\mu_1^2\mu_3\nu^3 + 6\mu_1^3\mu_3\nu^2 + 344\mu_1^2\mu_4\nu^3 + 16\mu_1^3\mu_4\nu^2 + 198\mu_2^2\mu_3\nu^3 + 608\mu_2^2\mu_4\nu^3 + 48\mu_1^2\mu_2^2\mu_3\mu_4 + 102\mu_1\mu_2^2\mu_3\nu^2 + 81\mu_1^2\mu_2\mu_3\nu^2 + 12\mu_1^2\mu_2^2\mu_3\nu + 272\mu_1\mu_2^2\mu_4\nu^2 + 216\mu_1^2\mu_2\mu_4\nu^2 + 32\mu_1^2\mu_2^2\mu_4\nu + 396\mu_1^2\mu_3\mu_4\nu^2 + 792\mu_2^2\mu_3\mu_4\nu^2 + 24\mu_1^3\mu_2\mu_3\mu_4 + 414\mu_1\mu_2\mu_3\nu^3 + 6\mu_1^3\mu_2\mu_3\nu + 1304\mu_1\mu_2\mu_4\nu^3 + 16\mu_1^3\mu_2\mu_4\nu + 1716\mu_1\mu_3\mu_4\nu^3 + 24\mu_1^3\mu_3\mu_4\nu + 2604\mu_2\mu_3\mu_4\nu^3 + 1656\mu_1\mu_2\mu_3\mu_4\nu^2 + 408\mu_1\mu_2^2\mu_3\mu_4\nu + 324\mu_1^2\mu_2\mu_3\mu_4\nu).$

\bigskip

Their ratio is 

\bigskip

 $ t_{L\rightarrow L}/ t_{1\rightarrow1} = \frac{(4\mu_4 + \nu)*(3\mu_3 + 2\nu)*(2\mu_1^3\mu_2 + 2\mu_1^3\nu + 4\mu_1^2\mu_2^2 + 17\mu_1^2\mu_2\nu + 13\mu_1^2\nu^2 + 14\mu_1\mu_2^2\nu + 43\mu_1\mu_2\nu^2 + 33\mu_1\nu^3 + 16\mu_2^2\nu^2 + 42\mu_2\nu^3 + 36\nu^4)}
 {2\nu^4(4\mu_1^2 + 10\mu_1\mu_2 + 24\mu_1\nu + 6\mu_3\mu_1 + 4\mu_2^2 + 30\mu_2\nu + 12\mu_3\mu_2 + 36\nu^2 + 27\mu_3\nu)}.$
 
\bigskip  

Average waiting times $t_i$ for avalanche of size $i$ in case
$\mu_i=\delta/i$, where $\delta=const.$ and $t:=\delta/\nu$, for $N=5$,  are 

\begin{eqnarray}
t_1 &=& \frac{4 t^5 + 48 t^4 + 237 t^3 + 603 t^2 + 762 t + 360}{\nu t^2 (4 t^3 + 28 t^2 + 69 t + 60)}, \label{eq:wt1}\\
t_2 &=& \frac{4 t^5 + 48 t^4 + 237 t^3 + 603 t^2 + 762 t + 360}{ 2\nu t^2(4 t^2 + 16 t + 15)}, \\
t_3 &=& \frac{4 t^6 + 56 t^5 + 333 t^4 + 1077 t^3 + 1968 t^2 + 1884 t + 720}{ 2\nu t^2 (10 t^2 + 31 t + 18)},\\%
t_4 &=& \frac{4 t^7 + 60 t^6 + 389 t^5 + 1410 t^4 + 3045 t^3 + 3852 t^2 + 2604 t + 720}{8\nu t^2 (7 t^2 + 24 t + 18)}, \\
t_5 &=& \frac{4 t^7 + 60 t^6 + 389 t^5 + 1410 t^4 + 3045 t^3 + 3852 t^2 + 2604 t + 720}{8\nu t(7 t^2 + 24 t + 18)}. \label{eq:wt5}
\end{eqnarray}

The ratio of return times  for $N=7$ for three cases:

for $\mu=const.$ 

\bigskip

$t_{L\rightarrow L}/ t_{1\rightarrow1}= (5184000t^{16} + 90633600t^{15} + 734038560t^{14} + 3656624904t^{13} + 12543798852t^{12} + 31435490078t^{11} + 59579986661t^{10} + 87223274254t^9 + 99846813214t^8 + 89833419890t^7 + 63379753809t^6 + 34652851894t^5 + 14319281196t^4 + 4279417752t^3 + 859191840t^2 + 101520000t + 5184000)/(5184000t^{10} + 60393600t^9 + 306948960t^8 + 896350104t^7 + 1664901648t^6 + 2053477662t^5 + 1700206878t^4 + 930252240t^3 + 320428800t^2 + 62380800t + 5184000), $

\bigskip

for $\mu_i/\nu=\theta/i$ 

\bigskip

$ t_{L\rightarrow L}/ t_{1\rightarrow1}=(576t^{16} + 16800t^{15} + 229696t^{14} + 1956752t^{13} + 11645844t^{12} + 51472058t^{11} + 175326610t^{10} + 471411274t^9 + 1015867913t^8 + 1768373403t^7 + 2486683328t^6 + 2797983376t^5 + 2465006400t^4 + 1636404624t^3 + 767257920t^2 + 225504000t + 31104000)/(76032t^{10} + 1450368t^9 + 12336072t^8 + 61572600t^7 + 199652130t^6 + 439389384t^5 + 664690536t^4 + 682575840t^3 + 455457600t^2 + 178329600t + 31104000),$

\bigskip

and for $\mu=\sigma/i^2$ 

\bigskip

$ t_{L\rightarrow L}/ t_{1\rightarrow1}= (27000t^{16} + 1648350t^{15} + 46021545t^{14} + 781598610t^{13} + 9060806565t^{12} + 76286696592t^{11} + 484749056302t^{10} + 2385421175676t^9 + 9253317988496t^8 + 28615082281632t^7 + 70836261328608t^6 + 139636245477312t^5 + 215233793554176t^4 + 250177275371520t^3 + 205696375603200t^2 + 106205478912000t + 25798901760000)/(1032264000t^{10} + 28816738200t^9 + 362493838440t^8 + 2695926522960t^7 + 13095290178720t^6 + 43360671643200t^5 + 99111840724224t^4 + 154547465674752t^3 + 157566016143360t^2 + 95025954816000t + 25798901760000). $

\bigskip

The exact value of $\alpha^E_{4}$ for $N=10$:
$$\alpha^E_{4}= \frac{2v_{20}+v_{43}+2v_{44}+2v_{45}+v_{46}+v_{72}+2v_{73}+v_{74}+v_{77}+2v_{79}+v_{95}+v_{96}}
{
2(v_{20}+v_{43}+v_{44}+v_{45}+v_{46}+v_{72}+v_{73}+v_{74}+v_{77}+v_{78}+v_{79}+v_{95}+v_{96}+v_{98}+v_{99}+2v_{106} )
}.$$

A prime factorization of the common factor of probabilities of states of FRDA for $N=10$ (presented in Table \ref{tab:statesN10})
for rebound parameters $\mu=1$ and $\nu=1$:
$$2^2*3^2*607*66617*3 56222 1887802308 6289926346 2293218927 8661478627 7415883867.$$
The biggest prime has 56 digits.


\begin{table*}[p]%
\begin{tabular}{c|c|c|c}
& states &  probability  & sym \\
\hline\hline
1 & $ | \ \ \ | \ \ \ | \ \ \ | \ \ \ | \ \ \ | \ \ \ | \ \ \ | \ \ \ | \ \ \ | \ \ \ | $	& 0.05402& $^{(1)}$ \\
\hline\hline
2 & $ | \ \ \ | \ \ \ | \ \ \ | \ \ \ | \ \ \ | \ \ \ | \ \ \ | \ \ \ | \ \ \ | \bullet | $	&0.13914 & \\
\hline\hline
3 & $ | \ \ \ | \ \ \ | \ \ \ | \ \ \ | \ \ \ | \ \ \ | \ \ \ | \ \ \ | \bullet | \bullet | $	&0.06060 &  \\
\hline
4 & $ | \ \ \ | \ \ \ | \ \ \ | \ \ \ | \ \ \ | \ \ \ | \ \ \ | \bullet | \ \ \ | \bullet | $	&0.04841 & \\
\hline
5 & $ | \ \ \ | \ \ \ | \ \ \ | \ \ \ | \ \ \ | \ \ \ | \bullet | \ \ \ | \ \ \ | \bullet | $	&0.04075 & \\
\hline
6 & $ | \ \ \ | \ \ \ | \ \ \ | \ \ \ | \ \ \ | \bullet | \ \ \ | \ \ \ | \ \ \ | \bullet | $	& 0.03777& \\
\hline
7 & $ | \ \ \ | \ \ \ | \ \ \ | \ \ \ | \bullet | \ \ \ | \ \ \ | \ \ \ | \ \ \ | \bullet | $	&0.01853 & $^{(5)}$ \\
\hline\hline
8 & $ | \ \ \ | \ \ \ | \ \ \ | \ \ \ | \ \ \ | \ \ \ | \ \ \ | \bullet | \bullet | \bullet | $	&0.03074 & \\
\hline
9 & $ | \ \ \ | \ \ \ | \ \ \ | \ \ \ | \ \ \ | \ \ \ |  \bullet| \ \ \ | \bullet | \bullet | $	&0.02256 & 14 \\
\hline
10 & $ | \ \ \ | \ \ \ | \ \ \ | \ \ \ | \ \ \ | \bullet | \ \ \ | \ \ \ | \bullet | \bullet | $	&0.01807 & 13 \\
\hline  
11 & $ | \ \ \ | \ \ \ | \ \ \ | \ \ \ | \bullet | \ \ \ | \ \ \ | \ \ \ | \bullet | \bullet | $	&0.01657 & 12 \\
\hline
12 & $ | \ \ \ | \ \ \ | \ \ \ | \bullet | \ \ \ | \ \ \ | \ \ \ | \ \ \ | \bullet | \bullet | $	&0.01657 & 11\\
\hline
13 & $ | \ \ \ | \ \ \ | \bullet | \ \ \ | \ \ \ | \ \ \ | \ \ \ | \ \ \ | \bullet | \bullet | $	& 0.01807& 10 \\
\hline
14 & $ | \ \ \ | \bullet | \ \ \ | \ \ \ | \ \ \ | \ \ \ | \ \ \ | \ \ \ | \bullet | \bullet | $	&0.02256 & 9 \\
\hline
15 & $ | \ \ \ | \ \ \ | \ \ \ | \ \ \ | \ \ \ | \bullet | \ \ \ | \bullet | \ \ \ | \bullet | $	&0.01744 & \\
\hline
16 & $ | \ \ \ | \ \ \ | \ \ \ | \ \ \ | \bullet | \ \ \ | \ \ \ | \bullet | \ \ \ | \bullet | $	&0.01450 & 18 \\
\hline
17 & $ | \ \ \ | \ \ \ | \ \ \ | \bullet | \ \ \ | \ \ \ | \ \ \ | \bullet | \ \ \ | \bullet | $	&0.01371 & \\
\hline
18 & $ | \ \ \ | \ \ \ | \bullet | \ \ \ | \ \ \ | \ \ \ | \ \ \ | \bullet | \ \ \ | \bullet | $	&0.01450 & 16 \\
\hline
19 & $ | \ \ \ | \ \ \ | \ \ \ | \bullet | \ \ \ | \ \ \ | \bullet | \ \ \ | \ \ \ | \bullet | $	& 0.01252 & \\
\hline\hline
20 & $ | \ \ \ | \ \ \ | \ \ \ | \ \ \ | \ \ \ | \ \ \ | \bullet | \bullet | \bullet | \bullet | $	&0.01641 & \\
\hline  
21 & $ | \ \ \ | \ \ \ | \ \ \ | \ \ \ | \ \ \ | \bullet | \ \ \ | \bullet | \bullet | \bullet | $	&0.01168 & 25 \\
\hline
22 & $ | \ \ \ | \ \ \ | \ \ \ | \ \ \ | \bullet | \ \ \ | \ \ \ | \bullet | \bullet | \bullet | $	&0.00926 & 24 \\
\hline
23 & $ | \ \ \ | \ \ \ | \ \ \ | \bullet | \ \ \ | \ \ \ | \ \ \ | \bullet | \bullet | \bullet | $	&0.00864 & \\
\hline
24 & $ | \ \ \ | \ \ \ | \bullet | \ \ \ | \ \ \ | \ \ \ | \ \ \ | \bullet | \bullet | \bullet | $	&0.00926 & 22 \\
\hline
25 & $ | \ \ \ | \bullet | \ \ \ | \ \ \ | \ \ \ | \ \ \ | \ \ \ | \bullet | \bullet | \bullet | $	&0.01168 & 21 \\
\hline
26 & $ | \ \ \ | \ \ \ | \ \ \ | \ \ \ | \ \ \ | \bullet |  \bullet| \ \ \ | \bullet | \bullet | $	&0.01075 & \\
\hline
27 & $ | \ \ \ | \ \ \ | \ \ \ | \ \ \ | \bullet | \bullet | \ \ \ | \ \ \ | \bullet | \bullet | $	& 0.00808 &\\
\hline
28 & $ | \ \ \ | \ \ \ | \ \ \ | \bullet | \bullet | \ \ \ | \ \ \ | \ \ \ | \bullet | \bullet | $	&0.00372 & $^{(5)}$ \\
\hline
29 & $ | \ \ \ | \ \ \ | \ \ \ | \ \ \ | \bullet | \ \ \ | \bullet | \ \ \ | \bullet | \bullet | $	& 0.00825 & 38\\
\hline
30 & $ | \ \ \ | \ \ \ | \ \ \ | \bullet | \ \ \ | \ \ \ | \bullet | \ \ \ | \bullet | \bullet | $	& 0.00697 & 37\\
\hline  
31 & $ | \ \ \ | \ \ \ | \bullet | \ \ \ | \ \ \ | \ \ \ | \bullet | \ \ \ | \bullet | \bullet | $	& 0.00703& 35 \\
\hline
32 & $ | \ \ \ | \bullet | \ \ \ | \ \ \ | \ \ \ | \ \ \ | \bullet | \ \ \ | \bullet | \bullet | $	& 0.00857& \\
\hline
33 & $ | \ \ \ | \ \ \ | \ \ \ | \bullet | \ \ \ | \bullet | \ \ \ | \ \ \ | \bullet | \bullet | $	& 0.00663& 36 \\
\hline
34 & $ | \ \ \ | \ \ \ | \bullet | \ \ \ | \ \ \ | \bullet | \ \ \ | \ \ \ | \bullet | \bullet | $	&0.00612 & \\
\hline
35 & $ | \ \ \ | \bullet | \ \ \ | \ \ \ | \ \ \ | \bullet | \ \ \ | \ \ \ | \bullet | \bullet | $	& 0.00703 & 31 \\
\hline
36 & $ | \ \ \ | \ \ \ | \bullet | \ \ \ | \bullet | \ \ \ | \ \ \ | \ \ \ | \bullet | \bullet | $	&0.00663 & 33 \\
\hline
37 & $ | \ \ \ | \bullet | \ \ \ | \ \ \ | \bullet | \ \ \ | \ \ \ | \ \ \ | \bullet | \bullet | $	&0.00697 & 30 \\
\hline
38 & $ | \ \ \ | \bullet | \ \ \ | \bullet | \ \ \ | \ \ \ | \ \ \ | \ \ \ | \bullet | \bullet | $	& 0.00825 & 29 \\
\hline
39 & $ | \ \ \ | \ \ \ | \ \ \ | \bullet | \ \ \ | \bullet | \ \ \ | \bullet | \ \ \ | \bullet | $	&0.00656  & \\
\hline
40 & $ | \ \ \ | \ \ \ | \bullet | \ \ \ | \ \ \ | \bullet | \ \ \ | \bullet | \ \ \ | \bullet | $	& 0.00590& \\
\hline
41 & $ | \ \ \ | \ \ \ | \bullet | \ \ \ | \bullet | \ \ \ | \ \ \ | \bullet | \ \ \ | \bullet | $	&0.00290 & $^{(5)}$ \\
\hline\hline   
42 & $ | \ \ \ | \ \ \ | \ \ \ | \ \ \ | \ \ \ | \bullet | \bullet | \bullet | \bullet | \bullet | $	& 0.00893& \\
\hline
43 & $ | \ \ \ | \ \ \ | \ \ \ | \ \ \ | \bullet | \ \ \ | \bullet | \bullet | \bullet | \bullet | $	&0.00632 & 46 \\
\hline
44 & $ | \ \ \ | \ \ \ | \ \ \ | \bullet | \ \ \ | \ \ \ | \bullet | \bullet | \bullet | \bullet | $	&0.00510 & 45\\
\hline
45 & $ | \ \ \ | \ \ \ | \bullet | \ \ \ | \ \ \ | \ \ \ | \bullet | \bullet | \bullet | \bullet | $	& 0.00510& 44 \\
\hline
46 & $ | \ \ \ | \bullet | \ \ \ | \ \ \ | \ \ \ | \ \ \ | \bullet | \bullet | \bullet | \bullet | $	&0.00632 & 43 \\
\hline
47 & $ | \ \ \ | \ \ \ | \ \ \ | \ \ \ | \bullet | \bullet | \ \ \ | \bullet | \bullet | \bullet | $	& 0.00563& 50 \\
\hline
48 & $ | \ \ \ | \ \ \ | \ \ \ | \bullet | \bullet | \ \ \ | \ \ \ | \bullet | \bullet | \bullet | $	&0.00424 & 49 \\
\hline
49 & $ | \ \ \ | \ \ \ | \bullet | \bullet | \ \ \ | \ \ \ | \ \ \ | \bullet | \bullet | \bullet | $	& 0.00424& 48 \\
\hline
50 & $ | \ \ \ | \bullet | \bullet | \ \ \ | \ \ \ | \ \ \ | \ \ \ | \bullet | \bullet | \bullet | $	&0.00563 & 47 \\
\hline
51 & $ | \ \ \ | \ \ \ | \ \ \ | \bullet | \ \ \ | \bullet | \ \ \ | \bullet | \bullet | \bullet | $	& 0.00441& 56 \\
\hline
52 & $ | \ \ \ | \ \ \ | \bullet | \ \ \ | \ \ \ | \bullet | \ \ \ | \bullet | \bullet | \bullet | $	&0.00399 & 55 \\
\hline
53 & $ | \ \ \ | \bullet | \ \ \ | \ \ \ | \ \ \ | \bullet | \ \ \ | \bullet | \bullet | \bullet | $	&0.00463 & \\
\hline
54 & $ | \ \ \ | \ \ \ | \bullet | \ \ \ | \bullet | \ \ \ | \ \ \ | \bullet | \bullet | \bullet | $	&0.00376 & \\
\hline
\end{tabular}
\quad \quad
\begin{tabular}{c|c|c|c}
& states &  probability & sym   \\
\hline\hline
55 & $ | \ \ \ | \bullet | \ \ \ | \ \ \ | \bullet | \ \ \ | \ \ \ | \bullet | \bullet | \bullet | $	&0.00399 & 52 \\
\hline
56 & $ | \ \ \ | \bullet | \ \ \ | \bullet | \ \ \ | \ \ \ | \ \ \ | \bullet | \bullet | \bullet | $	&0.00441 & 51 \\
\hline
57 & $ | \ \ \ | \ \ \ | \ \ \ | \bullet | \ \ \ | \bullet | \bullet | \ \ \ | \bullet | \bullet | $	& 0.00422 & 59 \\
\hline
58 & $ | \ \ \ | \ \ \ | \bullet | \ \ \ | \ \ \ | \bullet | \bullet | \ \ \ | \bullet | \bullet | $	&0.00370 & \\
\hline
59 & $ | \ \ \ | \bullet | \ \ \ | \ \ \ | \ \ \ | \bullet | \bullet | \ \ \ | \bullet | \bullet | $	& 0.00422 & 57 \\
\hline
60 & $ | \ \ \ | \ \ \ | \bullet | \ \ \ | \bullet | \bullet | \ \ \ | \ \ \ | \bullet | \bullet | $	& 0.00348& 61 \\
\hline
61 & $ | \ \ \ | \bullet | \ \ \ | \ \ \ | \bullet | \bullet | \ \ \ | \ \ \ | \bullet | \bullet | $	&0.00348 & 60 \\
\hline
62 & $ | \ \ \ | \bullet | \ \ \ | \bullet | \bullet | \ \ \ | \ \ \ | \ \ \ | \bullet | \bullet | $	&0.00401 & \\
\hline   
63 & $ | \ \ \ | \bullet | \ \ \ | \ \ \ | \bullet | \ \ \ | \bullet | \ \ \ | \bullet | \bullet | $	& 0.00355 & 66\\
\hline
64 & $ | \ \ \ | \ \ \ | \bullet | \ \ \ | \bullet | \ \ \ | \bullet | \ \ \ | \bullet | \bullet | $	& 0.00344& 65 \\
\hline
65 & $ | \ \ \ | \bullet | \ \ \ | \bullet | \ \ \ | \bullet | \ \ \ | \ \ \ | \bullet | \bullet | $	&0.00344 & 64 \\
\hline
66 & $ | \ \ \ | \bullet | \ \ \ | \bullet | \ \ \ | \ \ \ | \bullet | \ \ \ | \bullet | \bullet | $	&0.00355 & 63\\
\hline
67 & $ | \ \ \ | \bullet | \ \ \ | \bullet | \ \ \ | \bullet | \ \ \ | \bullet | \ \ \ | \bullet | $	& 0.00066& $^{(2)}$ \\
\hline\hline
68 & $ | \ \ \ | \ \ \ | \ \ \ | \ \ \ | \bullet | \bullet | \bullet | \bullet | \bullet | \bullet | $	&0.00492 & \\
\hline
69 & $ | \ \ \ | \ \ \ | \ \ \ | \bullet | \ \ \ | \bullet | \bullet | \bullet | \bullet | \bullet | $	& 0.00355 & 71\\
\hline
70 & $ | \ \ \ | \ \ \ | \bullet | \ \ \ | \ \ \ | \bullet | \bullet | \bullet | \bullet | \bullet | $	&0.00308 & \\
\hline
71 & $ | \ \ \ | \bullet | \ \ \ | \ \ \ | \ \ \ | \bullet | \bullet | \bullet | \bullet | \bullet | $	& 0.00355 & 69\\
\hline
72 & $ | \ \ \ | \ \ \ | \ \ \ | \bullet | \bullet | \ \ \ | \bullet | \bullet | \bullet | \bullet | $	& 0.00313& 74 \\
\hline
73 & $ | \ \ \ | \ \ \ | \bullet | \bullet | \ \ \ | \ \ \ | \bullet | \bullet | \bullet | \bullet | $	& 0.00256& \\
\hline
74 & $ | \ \ \ | \bullet | \bullet | \ \ \ | \ \ \ | \ \ \ | \bullet | \bullet | \bullet | \bullet | $	& 0.00313& 72 \\
\hline
75 & $ | \ \ \ | \ \ \ | \ \ \ | \bullet | \bullet | \bullet | \ \ \ | \bullet | \bullet | \bullet | $	& 0.00303& \\
\hline
76 & $ | \ \ \ | \ \ \ | \bullet | \bullet | \bullet | \ \ \ | \ \ \ | \bullet | \bullet | \bullet | $	&0.00122 & $^{(5)}$ \\
\hline
77 & $ | \ \ \ | \ \ \ | \bullet | \ \ \ | \bullet | \ \ \ | \bullet | \bullet | \bullet | \bullet | $	& 0.00266& 79 \\
\hline
78 & $ | \ \ \ | \bullet | \ \ \ | \ \ \ | \bullet | \ \ \ | \bullet | \bullet | \bullet | \bullet | $	&0.00277 & \\
\hline
79 & $ | \ \ \ | \bullet | \ \ \ | \bullet | \ \ \ | \ \ \ | \bullet | \bullet | \bullet | \bullet | $	& 0.00266 & 77\\
\hline
80 & $ | \ \ \ | \ \ \ | \bullet | \ \ \ | \bullet |  \bullet| \ \ \ | \bullet | \bullet | \bullet | $	& 0.00246 & 85\\
\hline
81 & $ | \ \ \ | \bullet | \ \ \ | \ \ \ | \bullet | \bullet | \ \ \ | \bullet | \bullet | \bullet | $	&0.00249 & 84 \\
\hline 
82 & $ | \ \ \ | \bullet | \ \ \ | \bullet | \bullet | \ \ \ | \ \ \ | \bullet | \bullet | \bullet | $	&0.00236 & 83 \\
\hline
83 & $ | \ \ \ | \ \ \ | \bullet | \bullet | \ \ \ | \bullet | \ \ \ | \bullet | \bullet | \bullet | $	&0.00236 & 82 \\
\hline
84 & $ | \ \ \ | \bullet | \bullet | \ \ \ | \ \ \ | \bullet | \ \ \ | \bullet | \bullet | \bullet | $	&0.00249 & 81 \\
\hline
85 & $ | \ \ \ | \bullet | \bullet | \ \ \ | \bullet | \ \ \ | \ \ \ | \bullet | \bullet | \bullet | $	&0.00246 & 80 \\
\hline
86 & $ | \ \ \ | \ \ \ | \bullet | \bullet | \ \ \ | \bullet | \bullet | \ \ \ | \bullet | \bullet | $	&0.00228 & \\
\hline
87 & $ | \ \ \ | \bullet | \ \ \ | \bullet | \ \ \ | \bullet | \ \ \ | \bullet | \bullet | \bullet | $	&0.00236  & \\
\hline
88 & $ | \ \ \ | \bullet | \ \ \ | \bullet | \ \ \ | \bullet | \bullet | \ \ \ | \bullet | \bullet | $	&0.00224 &  \\
\hline
89 & $ | \ \ \ | \bullet | \ \ \ | \bullet | \bullet | \ \ \ | \bullet | \ \ \ | \bullet | \bullet | $	&0.00111 & $^{(5)}$\\
\hline\hline
90 & $ | \ \ \ | \ \ \ | \ \ \ | \bullet | \bullet | \bullet | \bullet | \bullet | \bullet | \bullet | $	&0.00279 & \\
\hline
91 & $ | \ \ \ | \ \ \ | \bullet | \ \ \ | \bullet | \bullet | \bullet | \bullet | \bullet | \bullet | $	& 0.00219 & 92 \\
\hline
92 & $ | \ \ \ | \bullet | \ \ \ | \ \ \ | \bullet | \bullet | \bullet | \bullet | \bullet | \bullet | $	& 0.00219 & 91 \\
\hline
93 & $ | \ \ \ | \ \ \ | \bullet | \bullet | \ \ \ | \bullet | \bullet | \bullet | \bullet | \bullet | $	&0.00195 & 94 \\
\hline
94 & $ | \ \ \ | \bullet | \bullet | \ \ \ | \ \ \ | \bullet | \bullet | \bullet | \bullet | \bullet | $	&0.00195 & 93  \\
\hline
95 & $ | \ \ \ | \ \ \ | \bullet | \bullet | \bullet | \ \ \ | \bullet | \bullet | \bullet | \bullet | $	& 0.00186& 96 \\
\hline
96 & $ | \ \ \ | \bullet | \bullet | \bullet | \ \ \ | \ \ \ | \bullet | \bullet | \bullet | \bullet | $	& 0.00186& 95 \\
\hline
97 & $ | \ \ \ | \bullet | \ \ \ | \bullet | \ \ \ | \bullet | \bullet | \bullet | \bullet | \bullet | $	& 0.00194&  \\
\hline
98 & $ | \ \ \ | \bullet | \ \ \ | \bullet | \bullet | \ \ \ | \bullet | \bullet | \bullet | \bullet | $	&0.00178 & 99 \\
\hline
99 & $ | \ \ \ | \bullet | \bullet | \ \ \ | \bullet | \ \ \ | \bullet | \bullet | \bullet | \bullet | $	&0.00178 & 98 \\
\hline  
100 & $ | \ \ \ | \bullet | \ \ \ | \bullet | \bullet | \bullet | \ \ \ | \bullet | \bullet | \bullet | $	&0.00174 & \\
\hline
101 & $ | \ \ \ | \bullet | \bullet | \ \ \ | \bullet | \bullet | \ \ \ | \bullet | \bullet | \bullet | $	&0.00167 & \\
\hline\hline
102 & $ | \ \ \ | \ \ \ | \bullet | \bullet | \bullet | \bullet | \bullet | \bullet | \bullet | \bullet | $	&0.00176 & \\
\hline
103 & $ | \ \ \ | \bullet | \ \ \ | \bullet | \bullet | \bullet | \bullet | \bullet | \bullet | \bullet | $	&0.00164 & \\
\hline
104 & $ | \ \ \ | \bullet | \bullet | \ \ \ | \bullet | \bullet | \bullet | \bullet | \bullet | \bullet | $	& 0.00152& \\
\hline
105 & $ | \ \ \ | \bullet | \bullet | \bullet | \ \ \ | \bullet | \bullet | \bullet | \bullet | \bullet | $	&0.00147 & \\
\hline
106 & $ | \ \ \ | \bullet | \bullet | \bullet | \bullet | \ \ \ | \bullet | \bullet | \bullet | \bullet | $	& 0.00073& $^{(5)}$ \\
\hline\hline
107 & $ | \ \ \ | \bullet | \bullet | \bullet | \bullet | \bullet | \bullet | \bullet | \bullet | \bullet | $	&0.00142 & \\
\hline\hline
108 & $ | \bullet | \bullet | \bullet | \bullet | \bullet | \bullet | \bullet | \bullet | \bullet | \bullet | $	& 0.00014& $^{(1)}$ \\
\hline
\end{tabular}
\caption{All states for the size of the lattice $N=10$. The probabilities are for the case $\mu=1$ and $\nu=1$. States which 
differ by mirror symmetry are indicated in the column with label 'sym'. Values in parenthesis indicate smaller than $10$ translational symmetry.}
\label{tab:statesN10}
\end{table*}

\section*{Acknowledgement}

The author would like to express his gratitude to Professors Zbigniew Czechowski, Adam Doliwa and Maciej Wojtkowski
for inspiring comments and discussions.

\end{widetext}


\end{document}